\documentclass[msmath,amssymb,aps,pra,twocolumn,epsfig,showpacs,bibliography,lengthcheck,superscriptaddress]{revtex4-2}

\usepackage{graphicx}% Include figure files
\usepackage{dcolumn}% Align table columns on decimal point
\usepackage{bm}% bold math
\usepackage{hyperref}% add hypertext capabilities
\usepackage{epstopdf}
\usepackage{subcaption}
\usepackage{caption}
\usepackage{footmisc}
\usepackage{amsmath}
\usepackage{color}
\usepackage{ulem}

\begin{document}
\title{Dead-zone-free  free-induction-decay alkali-metal atomic magnetometer}%

\author{X.-K. Wang}
\author{J. He}
\author{H.-H. Deng}
\author{W. Gao}
\affiliation{Department of Precision Machinery and Precision Instrumentation, Key Laboratory of Precision Scientific Instrumentation of Anhui Higher Education Institutes, University of Science and Technology of China, Hefei 230027, China}

\author{D. Sheng}
\email{dsheng@ustc.edu.cn}
\affiliation{Department of Precision Machinery and Precision Instrumentation, Key Laboratory of Precision Scientific Instrumentation of Anhui Higher Education Institutes, University of Science and Technology of China, Hefei 230027, China}
\affiliation{Hefei National Laboratory, University of Science and Technology of China, Hefei 230088, China}

\begin{abstract}
The detection dead zone is an important systematic limitation in scalar atomic magnetometers, constraining their practical utility. In this work, we demonstrate a sensitive dead-zone-free scalar magnetometer by integrating previously established techniques into a FID magnetometer based on Bell–Bloom optical pumping. The dead zone is eliminated by inserting a reflecting mirror within a multipass-cavity-assisted atomic cell, which folds the optical beam into orthogonal paths. Our analysis reveals that {inter-region cross-talk} is essential for interpreting the resulted experimental signals. The sensor exhibits oscillation signal amplitude variations within a factor of three across all orientations in three-dimensional space, and a field sensitivity better than 80~fT/Hz$^{1/2}$ over the full space. We further characterize the heading error in the geomagnetic field range, where experimental results agree with theoretical predictions within 0.7~nT. Additionally, the sensor can operate in a closed-loop mode by feeding back the real-time frequency extracted from the FID signal to modulate the pump beam, we characterize the magnetic field slew rate of the sensor. This work pave the way towards a sensitive vector FID magnetometer.
\end{abstract}

\maketitle
\section{Introduction}

Free-induction-decay (FID) magnetometers operate on the principle of measuring external magnetic fields through the determination of the oscillation frequency in the FID signal of the atomic polarization.  In certain types of FID magnetometers, the pump beam is shut down when the FID signal is probed, which not only helps to avoid the light shift effect from the pump beam, but also eliminate the noise contribution from the pump beam. They have wide applications in fields such as fundamental physics~\cite{kimball2017,abel2020}, geomagnetic field detections~\cite{nightingale2025}, and biomagnetism imaging~\cite{limes2020} in the unshield environment. With the implementation of multipass cells, FID magnetometers have been developed into highly sensitive and precise scalar magnetometers~\cite{lee2021,liu2025}. 

In practical applications, the sensitivity and precision of FID magnetometers are limited by several key factors. First, heading error arises because the measured atomic Larmor frequency depends on the sensor’s orientation relative to the bias magnetic field. Second, drifts and fluctuations in the external field affect the signal amplitude particularly for FID magnetometers based on Bell-Bloom optical pumping, where optimal signal response requires the optical-pumping modulation frequency to be resonant with the atomic Larmor frequency. Maintaining this resonance typically demands real-time and reliable determination of the FID oscillation frequency to enable active closed-loop control of the pump modulation. Third, the signal amplitude also varies with sensor orientation, giving rise to a detection dead zone, a range of sensor orientations over which the signal drops substantially compared to its optimal value. Among these problems, heading error has been extensively studied both theoretically and experimentally~\cite{lee2021,liu2025}. Moreover, fast and sensitive frequency-counting methods based on the Hilbert transform have been developed for FID magnetometers~\cite{gong2025}. In this work, we focus primarily on the detection dead-zone problem, demonstrate a sensitive dead-zone-free FID magnetometer, and present detailed studies of its operation modes and parameters.

Two primary approaches have been developed to address the detection dead zone in atomic magnetometers. The first, introduced in Ref.~\cite{benkish2010}, modulates the optical polarization between circular and linear states, enabling simultaneous operation in Bell–Bloom optical-pumping and coherent-population-trapping modes, whose signals correspond respectively to the first and second harmonics of the atomic Larmor frequency. The resulting magnetometer is dead-zone free because the dead zones of the two modes are complementary. A similar scheme has been applied to a FID magnetometer using elliptically polarized light~\cite{mehta2025}. The second approach employs multiple sensors oriented so that their individual dead zones do not overlap~\cite{ness1970}. Building on the latter concept, a more compact configuration was demonstrated in Ref.~\cite{yu2022} by placing a reflecting mirror inside the atomic cell to fold the optical path into two orthogonal beams. This passive approach is employed as it allows for sensor miniaturization while maintaining a simple signal-analysis protocol {that divides the cell into two independent regions}. However, as detailed in the following sections, when implemented in a Herriott-cavity-enhanced cell, the full interpretation of the experimental data requires including {inter-region cross-talk effects}.

Following this introduction, Sec.~II describes the sensor setup, Sec.~III focuses on the detection dead zones and field sensitivity of the sensor, Sec.~IV studies two other parameters of the this sensor: heading error and magnetic field slew rate, and Sec.~V concludes the paper.

\section{Experiment setup}
\begin{figure}[htp]
\centering
\includegraphics[width=3in]{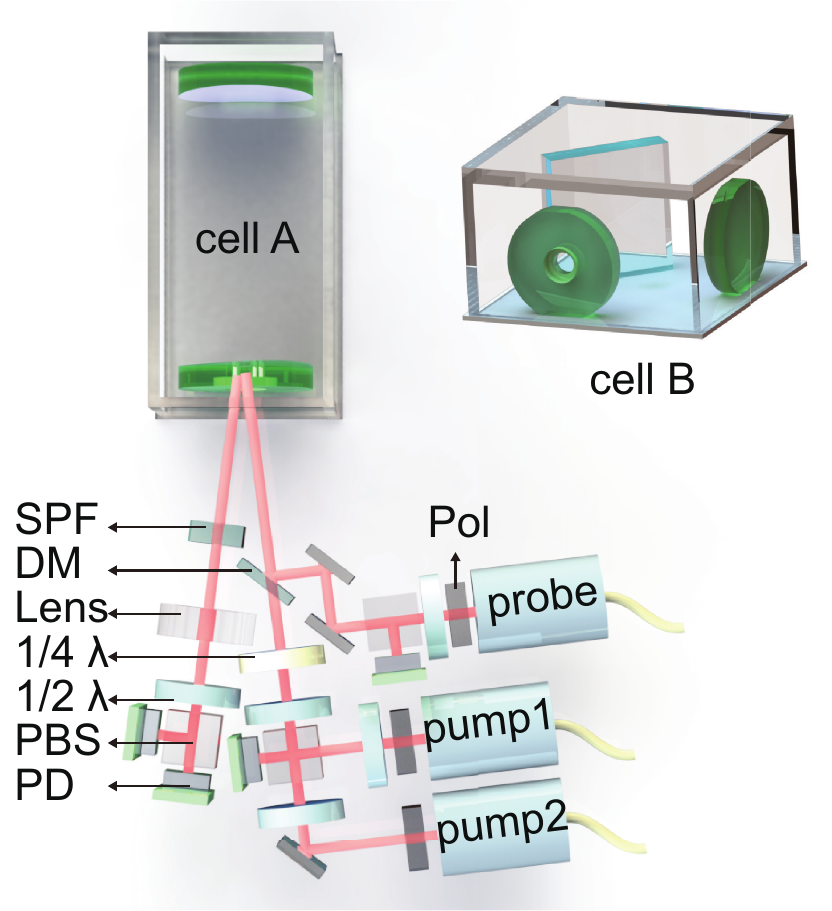}
\caption{\label{fig:setup} Optical configuration of the magnetometer sensor head. Two different Herriott-cavity-assisted cells have been used in this work, one is a two-mirror cell (cell A), and the other is a three-mirror cell with a reflecting mirror in the middle (cell B). SPF: short-pass filter, DM: dichroic mirror, PBS: polarizing beam splitter, PD: photodetector, Pol: polarizer, $\lambda/2$: half-wave plate, $\lambda/4$: quarter-wave plate.}
\end{figure}

The magnetometer sensor head is constructed on a 3D-printed optical platform, whose layout is illustrated in Fig.~\ref{fig:setup}. Pump and probe beams are generated by two independent distributed-Bragg-reflector (DBR) laser diodes. The pump beam is tuned to resonance with the Rb D1 line, whereas the probe beam is 120~GHz blue detuned from the Rb D2 line. Both beams are fiber-coupled into the sensor head. The pump beam can be delivered through either of two input channels, which produce output beams with orthogonal polarizations, and only one channel is active at a time. This dual-channel configuration enables the characterization of heading error.  Before entering the vapor cell, the pump and probe beams are first collimated with diameters of 1~mm, and then combined by a dichroic mirror. To improve the signal-to-noise ratio, we use Herriott-cavity-enhanced vapor cells, whose design will be detailed in the following section. After exiting the cell, the beams pass through a short-pass filter that strongly attenuates the pump light. The probe beam polarization is then analyzed by a polarimeter, consisting of a half-wave plate, a polarizing beam splitter, and two photodiode detectors. The subtraction of the two polarimeter photodiode signals constitutes the magnetometer output.

The magnetometer sensor is mounted inside a five-layer mu-metal magnetic shield assembly and connected via an epoxy holder to a high-precision rotation stage located outside the shields. Within the innermost shield, a plastic cylinder is positioned, with its surface wound with solenoid coils and cosine coils to generate the bias field inside the shielded volume. Additional gradient coils are implemented to compensate residual field gradients, which are maintained below 0.1~nT/mm during all measurements. 

%The $xyz$ coordinate system is established with reference to the cylindrical coil. The longitudinal axis of the cylinder is defined as the $z$-axis, while the $x$-axis is oriented along the horizontal direction within the transverse plane. 

\begin{figure}[htp]
\centering
\includegraphics[width=3in]{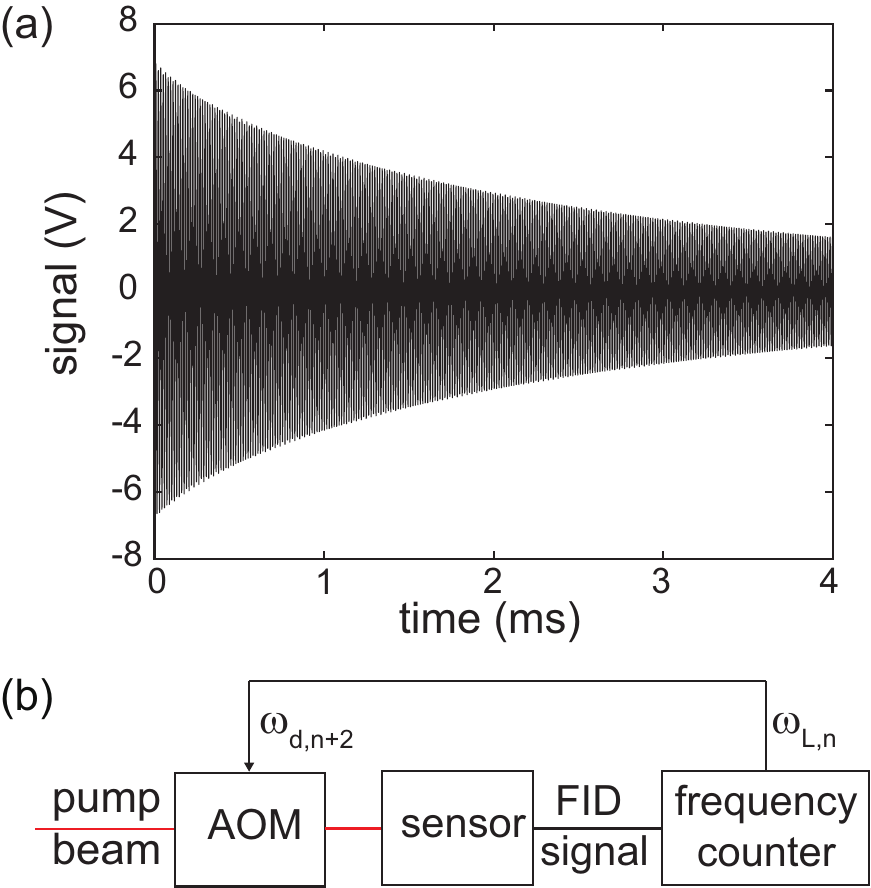}
\caption{\label{fig:sig} (a) A typical FID signals after passing through a high-pass filter with a bias field around 10~$\mu$T . (b) A schematic plot of the closed-loop operation mode, where $\omega_{L,n}$ obtained from the $n$th measurement cycle is used to set the AOM modulation frequency in the $(n+2)$th period.}
\end{figure}

The magnetometer operates in a pump–probe configuration. Each measurement cycle lasts 6~ms, comprising a 2~ms pumping interval followed by a 4~ms probing interval. During the pumping stage, the pump beam power is modulated via a fiber-coupled acousto-optic modulator (AOM) with a duty cycle of 20\%. When the modulation frequency $\omega_d$ is close to the atomic Larmor frequency $\omega_L$, this method can directly excite a sizable transverse atomic polarization~\cite{bell1961}. In the subsequent probe stage, the pump beam is switched off, and the FID signal is recorded. For an arbitrary sensor head orientation, the FID signal contains both a dc component and an ac component~\cite{liu2025}, which are determined by the longitudinal and transverse atomic polarization, respectively. We can separate the ac (dc) component of the FID signals by passing the magnetometer signal through a high-pass (low-pass) filter, and a typical ac signal is displayed in Fig.~\ref{fig:sig}(a).

In the absence of active bias-field adjustments, FID magnetometer operates stably in an open-loop mode under shielded environment. However, during the measurements of detection dead zones, the bias field direction needs to be periodically modified using different current combinations applied to the field coils, and this inevitably changes the magnitude of the total field as well. In this case, closed-loop operation is required to keep the pump-beam modulation frequency locked to $\omega_L$. As illustrated in Fig.~\ref{fig:sig}(b), we employ the frequency counter described in Ref.~\cite{gong2025} to analyze the whole signal from the probe interval, and feed the computed value $\omega_L$ back to the fiber AOM that drives the pump-beam power. It should be noted that, because of the finite processing time of the frequency counter, the feedback loop has a one-cycle delay: the result  $\omega_{L,n}$ obtained from the $n$th measurement cycle is used to set the AOM modulation frequency in the $(n+2)$th period. Although this delay is acceptable for dead‑zone measurements, it limits the magnetometer’s response time. To overcome this limitation, we will introduce a modified feedback strategy in Sec.~\ref{sec:hs}.

\section{Detection dead zones and magnetic field sensitivity}

{In this section, we first describe the experimental schemes used to measure the detection dead zones, presenting results for both cell A and B  illustrated in Fig.~\ref{fig:setup}. We then examine the field-orientation dependence of the atomic signals from cell B, and compare the independent-region model with a modified model that incorporates inter-region cross-talk. Finally, we characterize the full-space magnetic field sensitivity of the sensor based on cell B.}

\subsection{Measurement schemes}
\begin{figure}[htp]
\centering
\includegraphics[width=3in]{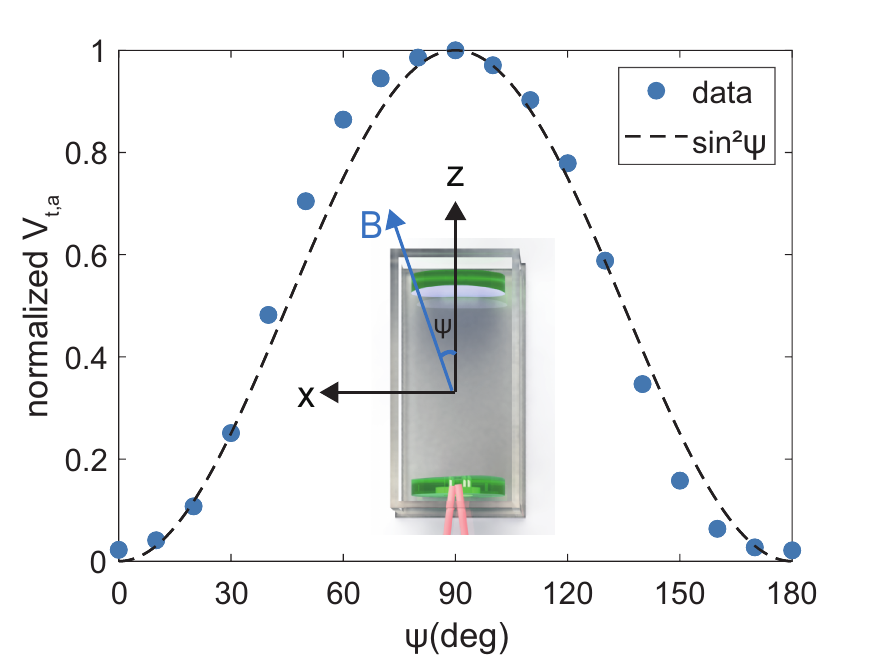}
\caption{\label{fig:dz}  {Experiment data (solid points) and theoretical predictions (dashed line) of normalized $V_{t,a}$ as a function of $\psi$ for a sensor based on cell A.}}
\end{figure}

The conventional Herriott-type cell, labeled cell A in Fig.~\ref{fig:setup} {with an inner dimension of 33 $\times$ 15 $\times$ 15~mm$^3$}, serves as the starting point for our design. The cell employs two cylindrical mirrors with identical specifications: a radius of curvature of 60 mm, a diameter of 12.7 mm, and a thickness of 2.1 mm. These mirrors are arranged with their principal axes separated by an angle of $51.28^\circ$ and are spaced 26.3 mm apart. The laser beam enters the cavity through a 2.5-mm-diameter through-hole at the center of the front mirror and exits from the same aperture after completing 14 passes inside the cavity. {The cell is filled with 350 torr N$_2$, and a droplet of $^{87}$Rb.} We performed measurements using this cell with a bias magnetic field of 8~$\mu$T. The measurement results {in Fig.~\ref{fig:dz}} show that the ac signal amplitude $V_{t,a}$ scales as $\sin^2\psi$, where $\psi$ is the angle between the applied bias field and the symmetry axis of the Herriott cavity, defined as the line connecting the centers of the two mirrors. This $\sin^2\psi$ dependence is in agreement with theoretical predictions~\cite{lee2021,liu2025} and results in detection dead zones near $\psi = 0$ and $\psi = \pi$, {where both the transverse polarization and its projection on the probe beam are small}.

\begin{figure}[htp]
\centering
\includegraphics[width=3in]{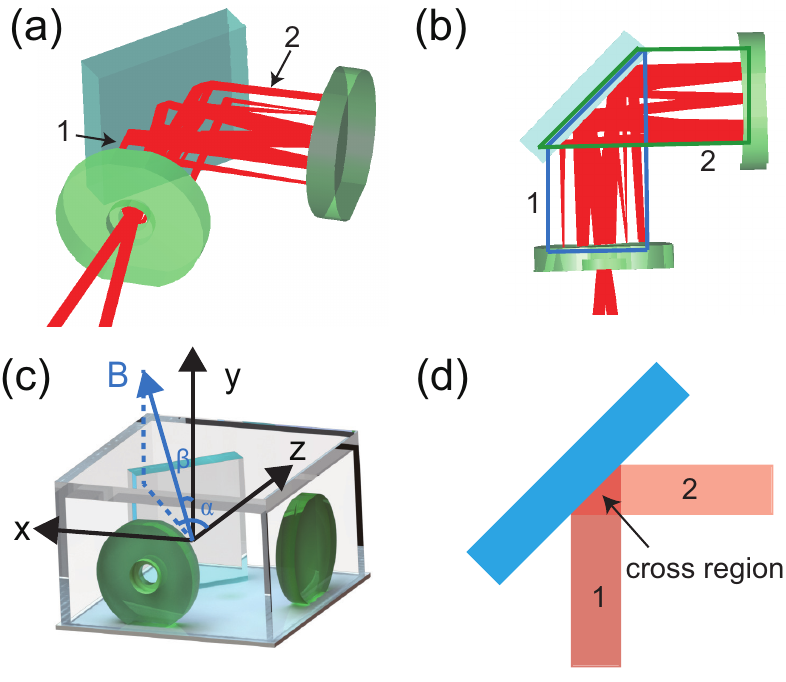}
\caption{\label{fig:3m} Plot (a) and (b) shows the 3D and 2D view of optical paths inside the three-mirror Herriott cavity, with a simple model has been proposed to divide the optical paths into two independent regions (1 and 2), {where beam directions and pump-beam polarizations are different in two regions due to the reflections by the middle mirror.} (c) Orientation of the bias field in the $xyz$ coordinate system fixed with the cell.  (d) An illustration of one beam path near the middle mirror.}
\end{figure}

To circumvent the detection dead zones in the configuration described above, we implemented the design demonstrated in Ref.~\cite{yu2022}, incorporating an additional mirror in the middle of the cavity. This modified configuration, referred to as cell B, is illustrated in the inset of Fig.~\ref{fig:setup}, {which has an inner dimension of 26 $\times$ 26 $\times$ 15~mm$^3$ and  contains the same gas and alkali atoms as the aforementioned cell A}. The introduction of this third mirror alters the intracavity optical path, generating two effectively orthogonal beam segments with complementary dead-zone orientations. In Ref.~\cite{yu2022}, a simplified model is proposed that divides the optical path into two distinct regions (labeled region 1 and region 2 in Fig.~\ref{fig:3m}(a) and (b)) based on the position of the extra mirror. This model treats the atomic ensembles sampled by these two regions as independent, effectively neglecting atomic diffusion between them. While this framework provides a useful qualitative understanding of the measurement outcomes, a quantitative verification has not been performed. In this work, we focus on a detailed comparison between this model and experimental data. This analysis allows for a deeper investigation of atomic dynamics within this three-mirror cavity, and thereby provides a reliable model for estimating other systematic effects, such as the heading error.

The related experiment was performed with the sensor held in a fixed orientation while the bias field direction was systematically scanned. We define a cell-fixed $xyz$ coordinate system, with the two optical regions inside the cell lying along the $x$- and $z$-axes as illustrated in Fig.~\ref{fig:3m}(c). The sensor was aligned so that the $z$ axis is parallel to the longitudinal direction of the shields and the $y$ axis is parallel to the local vertical direction. The bias field direction was controlled by independently varying the currents in the longitudinal and transverse field coils. For an arbitrary field orientation, depicted in Fig.~\ref{fig:3m}(c), its direction is parameterized by two angles: $\beta$, the polar angle relative to the $y$-axis, and $\alpha$, the azimuthal angle of its projection onto the $xz$-plane measured from the $z$-axis. A total of 266 distinct field directions were sampled. Throughout the scan, the field magnitude was maintained around 8~$\mu$T with a maximum deviation of  0.35~$\mu$T. To ensure that the measured signal amplitude depended solely on the field direction and not on its magnitude, the magnetometer was operated in a closed-loop configuration, which actively locked the pump beam modulation frequency to the atomic Larmor frequency.

\begin{figure*}[htp]
\centering
\includegraphics[width=\textwidth]{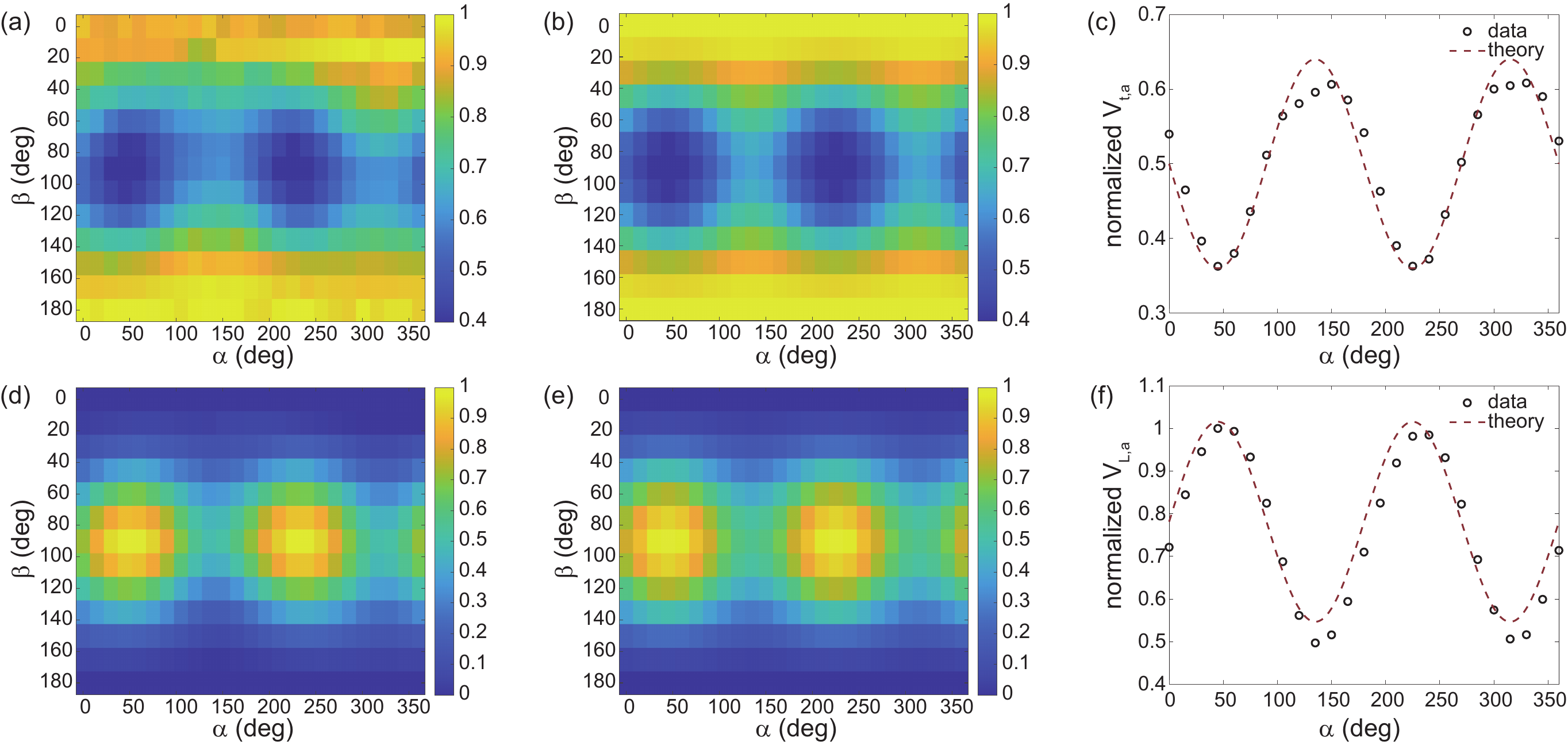}
\caption{\label{fig:theo} Plot (a) and (d) show normalized experiment results of the ac and dc signal amplitudes for all scanned bias field directions, respectively. Plot (b) and (e) show theoretical results based on Eq.~\eqref{eq:vta} and Eq.~\eqref{eq:vla} with $\eta=0.28$, respectively. Plot (c) and (f) show the comparison of the experiment and theoretical results when $\beta$ = $\pi/2$ for ac and dc signals, respectively. }
\end{figure*}

In the measurement, the cell temperature is stabilized at 70~$^\circ$C, the peak pump beam power is 5.5 mW before entering the cell, the transmitted probe beam power is 2 mW, and the atomic polarization is estimated to be 75\% using the methods in Ref.~\cite{liu2025}. The FID signal is recorded with a sampling rate of 1 MHz, and the amplitudes of its ac and dc components, normalized by their corresponding maximum observed values over the scanned angular space, are plotted in Figs.~\ref{fig:theo}(a) and (d), respectively. The maximum ac signal amplitude occurs when $\beta = 0$ or $\pi$, where the dc signal amplitude shows minimum values. The minimum ac signal amplitude, approximately 35\% of its maximum value at $\beta = 0$, occurs when $\beta = \pi/2$ and $\alpha$ is around $\pi/4$ or $5\pi/4$, where the dc signal amplitude shows maximum values.

\subsection{Theoretical models of atomic signals in cell B}
Following the aforementioned independent-region model, we analyze the atomic responses in each region separately. For the atoms in region $i$, under the high-polarization limit, the atomic polarization $\mathbf{P}_i$ is aligned with the photon spin $\hat{s}_i$ of the pump beam~\cite{appelt98}. The transverse ($\mathbf{P}_{i,t}$) and longitudinal ($\mathbf{P}_{i,L}$) components of the atomic polarization in region $i$ can be expressed as
\begin{eqnarray}
\mathbf{P}_{i,t}&=&P[\mathbf{a}_i\cos(\omega t+\phi)+\mathbf{b}_i\sin(\omega t+\phi)],\\
\mathbf{P}_{i,L}&=&P(\hat{s}_i\cdot\hat{B})\hat{B},\\
\mathbf{a}_i&=&\hat{s}_i-(\hat{s}_i\cdot\hat{B})\hat{B},\\
\mathbf{b}_i&=&\mathbf{a}_i\times\hat{B}=\hat{s}_i\times{\hat{B}},
\end{eqnarray}
where $\hat{B}=(\sin\beta\sin\alpha, \cos{\beta}, \sin\beta\cos\alpha)$ is the unit vector along the bias field direction. In the probe stage, the measured signal is determined by the projection of the atomic polarization components onto the beam propagation direction. Therefore, the signal amplitudes follow the relations:
\begin{eqnarray}
V_{1,t}&\propto&\mathbf{P}_{1,t}\cdot\hat{k}_1=Ps_1(1-\sin^2\beta\cos^2\alpha)\cos(\omega t+\phi),\\
V_{2,t}&\propto&\mathbf{P}_{2,t}\cdot\hat{k}_2=-Ps_2(1-\sin^2\beta\sin^2\alpha)\cos(\omega t+\phi),\\
V_{1,L}&\propto&\mathbf{P}_{1,L}\cdot\hat{k}_1=Ps_1\sin^2\beta\cos^2\alpha,\\
V_{2,L}&\propto&\mathbf{P}_{2,L}\cdot\hat{k}_2=-Ps_2\sin^2\beta\sin^2\alpha,
\end{eqnarray}
where $\hat{s}_1 = s_1 \hat{z}$, $\hat{s}_2 = s_2 \hat{x}$, $\phi$ is a phase delay, and $\hat{k}_i$ is the unit vector along the propagation direction of the pump beam in region $i$.

In this work, the reflecting mirror in the middle of the cavity is coated with dielectric materials, which introduces a $\pi$ phase shift between the $s$ and $p$ polarization components of the beam upon reflection. This results in $s_1 = s_2$, and it can also be shown that the resulting signals for the transverse and longitudinal polarizations are $V_{t} = V_{1,t} - V_{2,t}$ and $V_{L} = V_{1,L} - V_{2,L}$, respectively. Therefore, the corresponding signal amplitudes are given by these combinations:
\begin{eqnarray}
V_{t,a}&\propto& P(2-\sin^2\beta),~\label{eq:indpt}\\
V_{L,a}&\propto& P\sin^2\beta.
\end{eqnarray}
From the equations above, it can be concluded that the independent-region model predicts that both ac and dc signal amplitudes depend only on $\beta$, which is in clear contradiction with the experimental data plotted in Figs.~\ref{fig:theo}(a) and (d).
 
{The failure of the independent-region model reflects its oversimplified treatment of the optical paths inside the cavity. Several mechanisms may contribute to the deviation between the experimental data and Eq.~\eqref{eq:indpt}. Unequal path lengths in regions 1 and 2, caused by a displacement of the middle mirror from its designed position, could in principle contribute to this deviation. However, given the 0.1~mm fabrication tolerance of the cavity, this contribution is expected to be minor compared with the observed effect.} 
 
{More relevant than this fabrication error are the local effects of optical-field overlap and atomic diffusion near the middle  mirror. In the overlap region between the incoming and reflected beams (see Fig.~\ref{fig:3m}(d)), the higher-intensity beam predominantly polarizes the atoms. These polarized atoms can then be detected by the probe beams in both regions, giving rise to cross-talk effect: atoms initially polarized by the pump beam in one region are detected by the probe beam in the other. Geometrically, the magnitude of this effect, which is determined by the ratio of the beam size over the cavity length, is on the order of 10\% in our setup. Atomic diffusion can further enhance this cross-talk by transporting atoms from beam paths in one region into beam paths in the other region during the probe time.  With a diffusion constant $D \approx$ 0.33~cm$^2$/s for Rb atoms in 350 Torr N$_2$ buffer gas,  the characteristic two-dimensional diffusion length $L_D = 2\sqrt{DT_2} \sim$ 0.6~mm is well below the minimum  spacing (1.6~mm) between adjacent spots in the middle of the Herriott cell (14 spots per transverse plane), so inter-spot diffusion is negligible. The only relevant diffusive process is therefore the local mixing at each spot, which transports atoms from the non-overlapping region of a beam path into its overlapping region. }

To incorporate these cross-region effects into the modified model, we define a parameter $\eta$ to describe the signal contribution from polarized atoms diffusing in from the other region. The resulting ac signal measured by the probe beam in region 1 is then modified as follows:
\begin{eqnarray}~\label{eq:v1t'}
V'_{1,t}&\propto&\mathbf{P}_{1,t}\cdot\hat{k}_1+\eta\mathbf{P}_{2,t}\cdot\hat{k}_1\nonumber\\
&=&Ps_1\left[\left(1-\sin^2\beta\cos^2\alpha-\frac{\eta}{2}\sin^2\beta\sin2\alpha\right)\right.\nonumber\\
&&\left.\cos(\omega t+\phi)+\eta(\hat{k}_1\times\hat{x})\cdot\hat{B}\sin(\omega t+\phi)\right].
\end{eqnarray}
{The mixing due to the cross-talk effect exhibits a dependence on the field orientation, because the relative weight of the mixed component is field-orientation dependent, while the atomic diffusion process itself is still isotropic.}

Similarly, it can be shown that the ac signal measured by probe beams in region 2 is
\begin{eqnarray}~\label{eq:v2t'}
V'_{2,t}&\propto&\mathbf{P}_{2,t}\cdot\hat{k}_2+\eta\mathbf{P}_{1,t}\cdot\hat{k}_2\nonumber\\
&=&-Ps_1\left[\left(1-\sin^2\beta\sin^2\alpha-\frac{\eta}{2}\sin^2\beta\sin2\alpha\right)\right.\nonumber\\
&&\left.\cos(\omega t+\phi)-\eta(\hat{k}_2\times\hat{z})\cdot\hat{B}\sin(\omega t+\phi)\right].
\end{eqnarray}
Therefore, the total oscillation signal and its amplitude in the modified model can be expressed as
\begin{eqnarray}
V'_{t}&\propto&Ps_1\left(2-\sin^2\beta-\eta\sin^2\beta\sin2\alpha\right)\cos(\omega t+\phi),\\
V'_{t,a}&\propto&P\left(2-\sin^2\beta-\eta\sin^2\beta\sin2\alpha\right)~\label{eq:vta}.
\end{eqnarray}
In the same way, we can get expressions of the modified total longitudinal polarization signal and its amplitude as
\begin{eqnarray}
V'_{L}&\propto&Ps_1\sin^2\beta(1+\eta\sin2\alpha),\\
V'_{L,a}&\propto&P\sin^2\beta(1+\eta\sin2\alpha)~\label{eq:vla}.
\end{eqnarray}

Using Eqs.~\eqref{eq:vta} and~\eqref{eq:vla} to fit the experimental data in Figs.~\ref{fig:theo}(a) and (d), we obtain fitting results of $\eta = 0.24$ and $\eta = 0.32$, respectively. We take the average of these two values as $\eta = 0.28$ and plot the predicted normalized ac and dc signal amplitudes in Figs.~\ref{fig:theo}(b) and (e). A selective comparison between the theoretical and experimental results for $\beta = \pi/2$ is shown in Figs.~\ref{fig:theo}(c) and (f). Compared with the independent-region model, this modified model shows better agreement with the experimental results and {qualitatively} confirms the important role of inter-region cross-talk effects in this configuration. {A quantitative understanding of $\eta$ is under investigation}.

\subsection{Magnetic field sensitivity}
\begin{figure}[htp]
\centering
\includegraphics[width=3in]{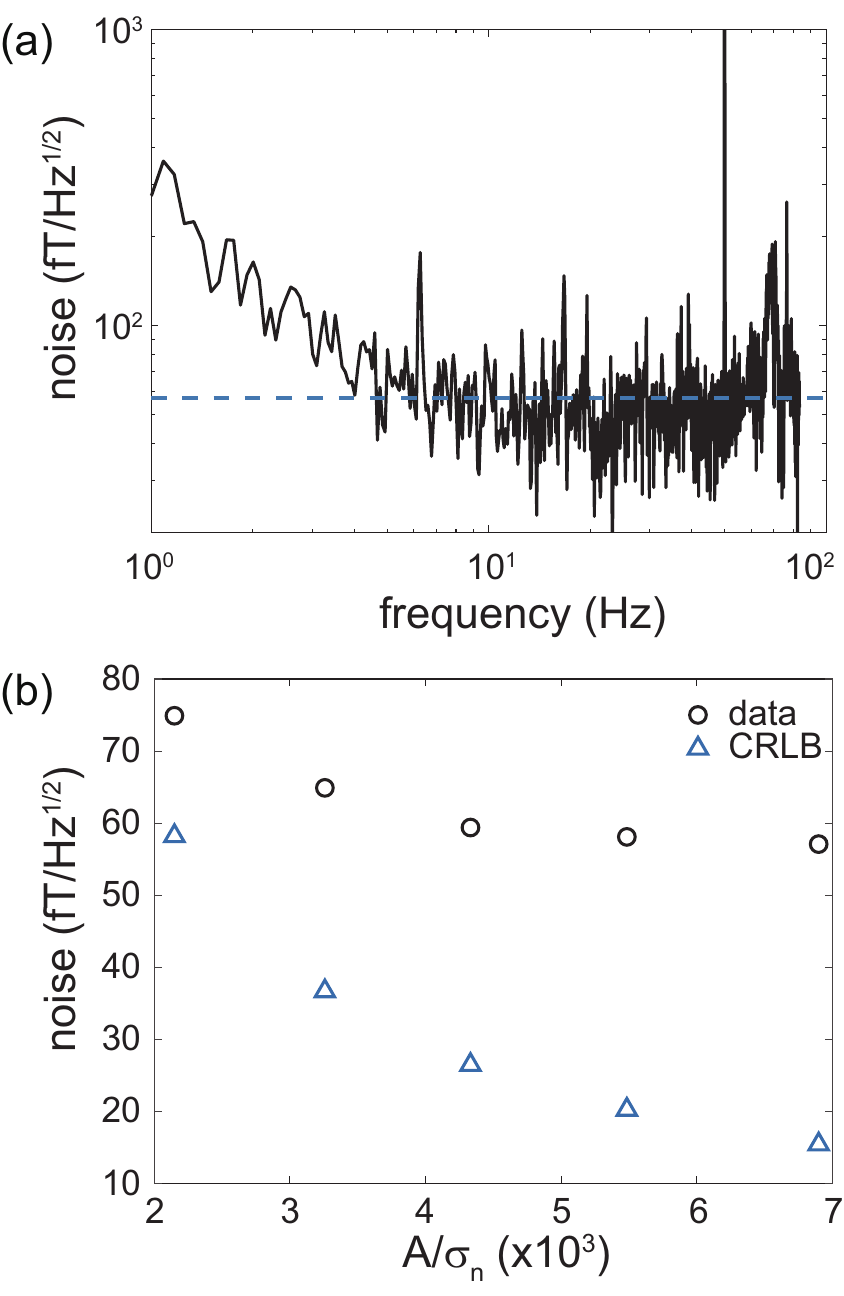}
\caption{\label{fig:sens}(a) Measured field noise when the bias field is 10 $\mu$T, $\beta = 0$, $A/\sigma_n= 6.9\times 10^3$, and $T_2= 2.6$~ms. The dash line denotes the measured field noise of 58 fT/Hz$^{1/2}$ when averaged over the frequency range from 5 to 15 Hz.  (b) Variation of calculated CRLB and measured field noise averaged over the frequency range from 5 to 15 Hz, as $A/\sigma_n$ changes from $6.9\times 10^3$ to $2.2 \times 10^3$.}
\end{figure}

To measure the magnetic field sensitivity, we set the magnitude of the bias field to approximately 10 $\mu$T and its direction along the $y$ axis in Fig.~\ref{fig:3m}(c), so that $\beta = 0$ and the ac signal amplitude from the atoms is maximized. Other experimental parameters are the same as those in the aforementioned measurements. In this case, the signal-to-noise ratio for the transverse atomic polarization in Eq.~\eqref{eq:vars} is $A/\sigma_n = 6.9 \times 10^3$, and the transverse depolarization rate is $1/T_2 = 385$~s$^{-1}$, {where the contribution from the atom-probe-beam interaction is 166~s$^{-1}$}. The magnitude of the magnetic field can be extracted from the oscillation frequencies of the FID signals, and its noise is plotted in Fig.~\ref{fig:sens}(a). The measured field noise is approximately 58 fT/Hz$^{1/2}$ when averaged over the frequency range from 5 to 15 Hz, where the CRLB of magnetometer noise is 16 fT/Hz$^{1/2}$ using Eq.~\eqref{eq:rhow}, and the current source noise is around 55 fT/Hz$^{1/2}$. We also attempted to reduce $A$ by decreasing the pump beam power and observed that the measured field noise remains within 80~fT/Hz$^{1/2}$  when $A/\sigma_n$ is lowered to $2.2 \times 10^3$ , which mimics the reduction of the signal-to-noise ratio from its maximum to its minimum in the three-dimensional space shown in Fig.~\ref{fig:theo}(a). Therefore, it can be concluded that the magnetic field sensitivity of the sensor is better than 80 fT/Hz$^{1/2}$ over the full space.
 
\section{Heading error and slew rate}~\label{sec:hs}

As demonstrated in Ref.~\cite{liu2025}, the heading error of FID magnetometers based on Bell-Bloom optical pumping is mainly due to the nonlinear Zeeman effect. In the high-polarization limit, the measured magnitude of the bias field using $^{87}$Rb atoms can be expressed as~\cite{lee2021,liu2025}
\begin{equation}~\label{eq:B87}
\mathrm{B}_m\approx\frac{4\hbar\omega_L}{(g_s-3g_I)\mu_B}\left[1+\frac{3\omega_L}{\omega_{hf}}\cos\psi\frac{P(7+P^2)}{5+3P^2}\right],
\end{equation}
where $\psi$ is the angle between the pump beam propagation direction and the bias field, $\omega_L$ is the measured Larmor frequency, $\hbar\omega_{hf}$ is the hyperfine splitting,  $\mu_B$ is the Bohr magneton, $g_s\approx2.002319$, and $g_I\approx0.000995$~\cite{liu2025} for $^{87}$Rb. The second term in the equation above represents the heading error
\begin{equation}
\Delta\mathrm{B}\approx \mathrm{B}_m\frac{3\omega_L}{\omega_{hf}}\cos\psi\frac{P(7+P^2)}{5+3P^2}=\Delta\mathrm{B}_0(P)\cos\psi,
\end{equation}
with $\Delta \mathrm{B}_0$=7.7 nT when $P=1$ and $\mathrm{B}_m= 50$~$\mu$T. {These heading-error results are valid under two conditions. First, the atomic polarization needs to be high. As demonstrated in Ref.~~\cite{lee2021}, when the atomic polarization exceeds 0.7, the discrepancy between predictions based on Eq.~(17) and full density-matrix calculations is within 5\%. Second, the atomic distribution needs to be close to the spin-temperature distribution, which has been confirmed to be true for Bell-Bloom optical pumping at a cell temperature around 70$^\circ$C in Ref.~\cite{liu2025}.}

\begin{figure}[htp]
\centering
\includegraphics[width=3in]{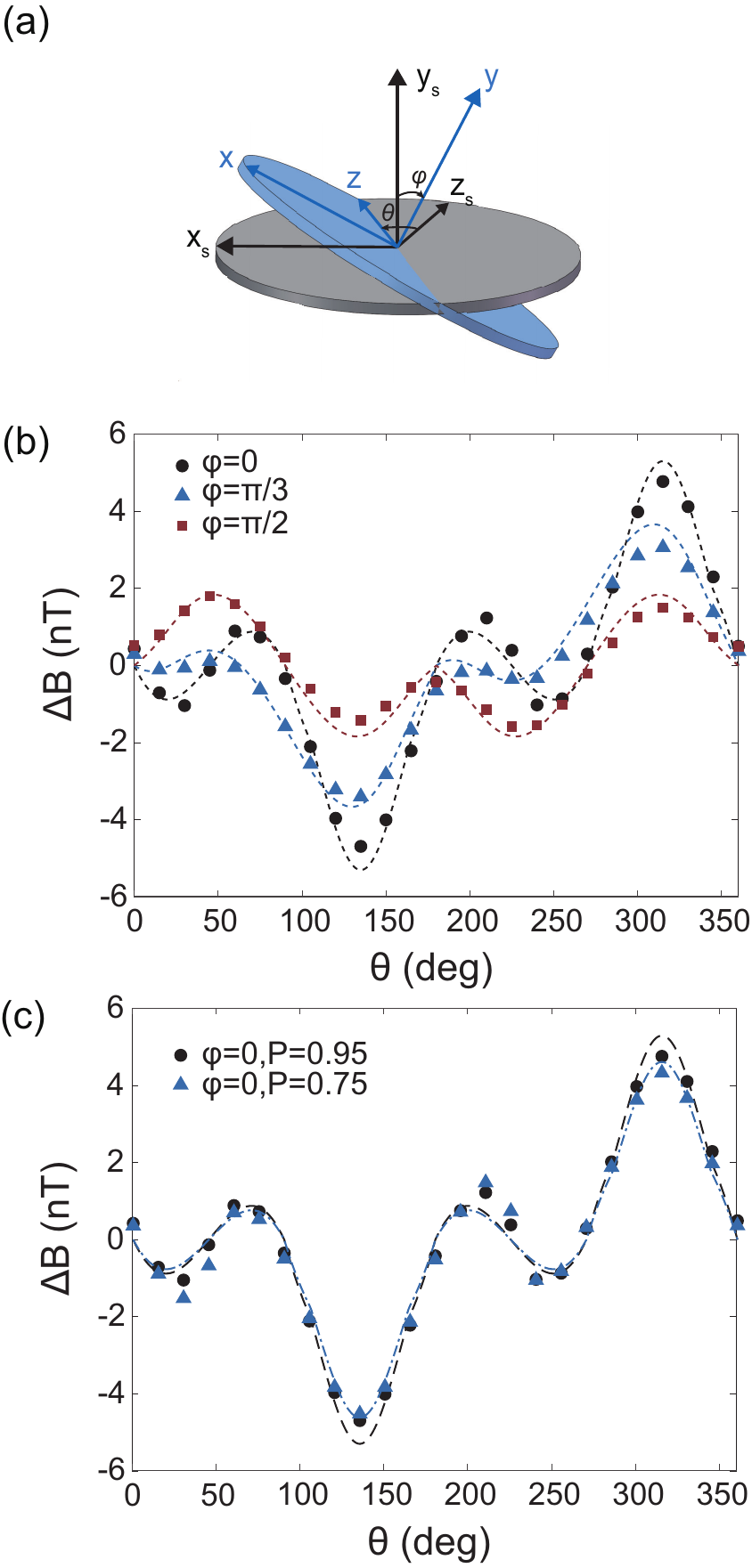}
\caption{\label{fig:herr}(a) Two coordinates used in the heading error measurement. The $x_sy_sz_s$ coordinate with $z_s$ along the bias field direction, and $y_s$ axis along the local vertical direction. The $xyz$ coordinate is the fixed with the atomic cell as shown in Fig.~\ref{fig:3m}(c). (b) Experiment (points) and theoretical results (dash lines) of heading error as a function of $\theta$ at $\varphi=0$, $\pi/3$, and $\pi/2$ with a bias field of 50~$\mu$T.  {(c) Experiment (points) and theoretical (dash lines) results of heading error as a function of $\theta$ at $\varphi=0$ with different atomic polarization.}}
\end{figure}

To measure the heading error of the detection-dead-zone-free sensor described in the previous section, we fix the bias field direction along the longitudinal axis of the shields and vary the angle $\psi$ by rotating the sensor head via the rotation table connected to it outside the shields. To describe the measurement results, we define a second coordinate system $x_sy_sz_s$ in addition to the $xyz$ coordinate system which is fixed to the vapor cell as shown in Fig.~\ref{fig:3m}(c). This $x_s y_s z_s$ system is fixed to the shields, with the $z_s$ axis along the bias field direction, and the $x_s$ and $y_s$ axes along the local horizontal and vertical directions, respectively.  As shown in Fig.~\ref{fig:herr}(a), a vector $\mathbf{V}$ in the $xyz$ coordinate system is related to its counterpart $\mathbf{V}_s$ in the $x_s y_s z_s$ system by the transformation
\begin{equation}
V=R_{ys}(\theta)R_{zs}(\varphi)V_s,
\end{equation}
where $R_i(\phi)$ represents a rotation about the $i$ axis by an angle $\phi$. Consequently, the parameters $\alpha$ and $\beta$ in Fig.~\ref{fig:3m}(c) are related to $\theta$ and $\varphi$ in Fig.~\ref{fig:herr}(a) by the following relations:
\begin{eqnarray}~\label{eq:co}
\sin\beta\sin\alpha&=&-\sin\theta\cos\varphi,\nonumber\\
\sin\beta\cos\alpha&=&\cos\theta,\nonumber\\
\cos\beta&=&\sin\theta\sin\varphi.
\end{eqnarray}    
  
Combining Eqs.~\eqref{eq:v1t'},~\eqref{eq:v2t'}, and~\eqref{eq:co}, we obtain the heading error ($\Delta B_1$) for atoms pumped by beams in region 1 of Fig.~\ref{fig:3m}(a), the corresponding measured ac signal ($V_{B1,t}$), and its amplitude ($V_{B1,ta}$) as
\begin{eqnarray}
\Delta \mathrm{B}_1&=&\Delta \mathrm{B}_0(P)\cos\theta\\
V_{B1,t}&\propto&\mathbf{P}_{1,t}\cdot\hat{k}_1-\eta\mathbf{P}_{1,t}\cdot\hat{k}_2\nonumber\\
&=&Ps_1\left[\left(1-\cos^2\theta+\frac{\eta}{2}\cos\varphi\sin2\theta\right)\right.\nonumber\\
&&\left.\cos(\omega t+\phi)+\eta\sin\theta\sin\varphi\sin(\omega t+\phi)\right],\\
V_{B1,ta}&\propto&P\left[\left(1-\cos^2\theta+\frac{\eta}{2}\cos\varphi\sin2\theta\right)^2+\right.\nonumber\\
&&\left.\eta^2\sin^2\theta\sin^2\varphi\right]^{1/2}.
\end{eqnarray} 
Similarly, for atoms pumped by beams in region 2 of Fig.~\ref{fig:3m}(a), we obtain
\begin{eqnarray}
\Delta \mathrm{B}_2&=&-\Delta \mathrm{B}_0(P)\sin\theta\cos\varphi,\\
V_{B2,t}&\propto&-\mathbf{P}_{2,t}\cdot\hat{k}_2+\eta\mathbf{P}_{2,t}\cdot\hat{k}_1\nonumber\\
&=&Ps_1\left[\left(1-\sin^2\beta\sin^2\alpha+\frac{\eta}{2}\cos\varphi\sin2\theta\right)\right.\nonumber\\
&&\left.\cos(\omega t+\phi)-\eta\sin\theta\sin\varphi\sin(\omega t+\phi)\right],\\
V_{B2,ta}&\propto&P\left[\left(1-\sin^2\beta\sin^2\alpha+\frac{\eta}{2}\cos\varphi\sin2\theta\right)^2+\right.\nonumber\\
&&\left.\eta^2\sin^2\theta\sin^2\varphi\right]^{1/2}.
\end{eqnarray}
The resulting heading error is the average of the heading errors from the equations above, weighted by their amplitudes:
\begin{equation}~\label{eq:DB}
\Delta \mathrm{B}=\frac{V_{B1,ta}\Delta\mathrm{B}_1+V_{B2,ta}\Delta\mathrm{B}_2}{V_{B1,ta}+V_{B2,ta}}.
\end{equation}

In the measurements, we set the bias field to 50 $\mu$T and increase the peak pump beam power to 15 mW, so that the atomic polarization is increased to 95\%. In addition, we performed back-to-back measurements in which the input port of the pump beam was switched between channel 1 and channel 2, as shown in Fig.~\ref{fig:setup}, so that the pump beam polarizations are opposite in consecutive measurements. For a given orientation of the sensor, let the magnetometer readout with pump beam polarization $\sigma+$ ($\sigma-$) be $\mathrm{B}_+$ ($\mathrm{B}_-$). Then the measured heading error at this orientation can be expressed as
\begin{equation}~\label{eq:avghe}
\Delta \mathrm{B}(\theta,\varphi)=\frac{\mathrm{B}_+(\theta,\varphi)-\mathrm{B}_-(\theta,\varphi)}{2}.
\end{equation}
Figure~\ref{fig:herr}(b) shows the experimental data for the extracted heading error using Eq.~\eqref{eq:avghe} at $\varphi = 0$, $\pi/3$, and $\pi/2$, together with the theoretical predictions using Eq.~\eqref{eq:DB}. The discrepancies between the experiment and theory are within 0.7~nT, {which is also true for most of the field orientations when atomic polarization is reduced to 0.75 (see Fig.~\ref{fig:herr}(c)).}

Another important magnetometer parameter is its detection bandwidth. The bandwidth is normally calibrated by an ac signal with an amplitude relatively small compared with the line width of the response curve of the magnetometer, and the result is half of the FID magnetometer repetition rate~\cite{yi2024}, limited by the Nyquist sampling theorem. The repetition-rate-limited bandwidth, or the cut-off signal frequency, of the magnetometer presented in Fig.~\ref{fig:sens}(a) is 83.3 Hz. 

\begin{figure}[htp]
\centering
\includegraphics[width=3in]{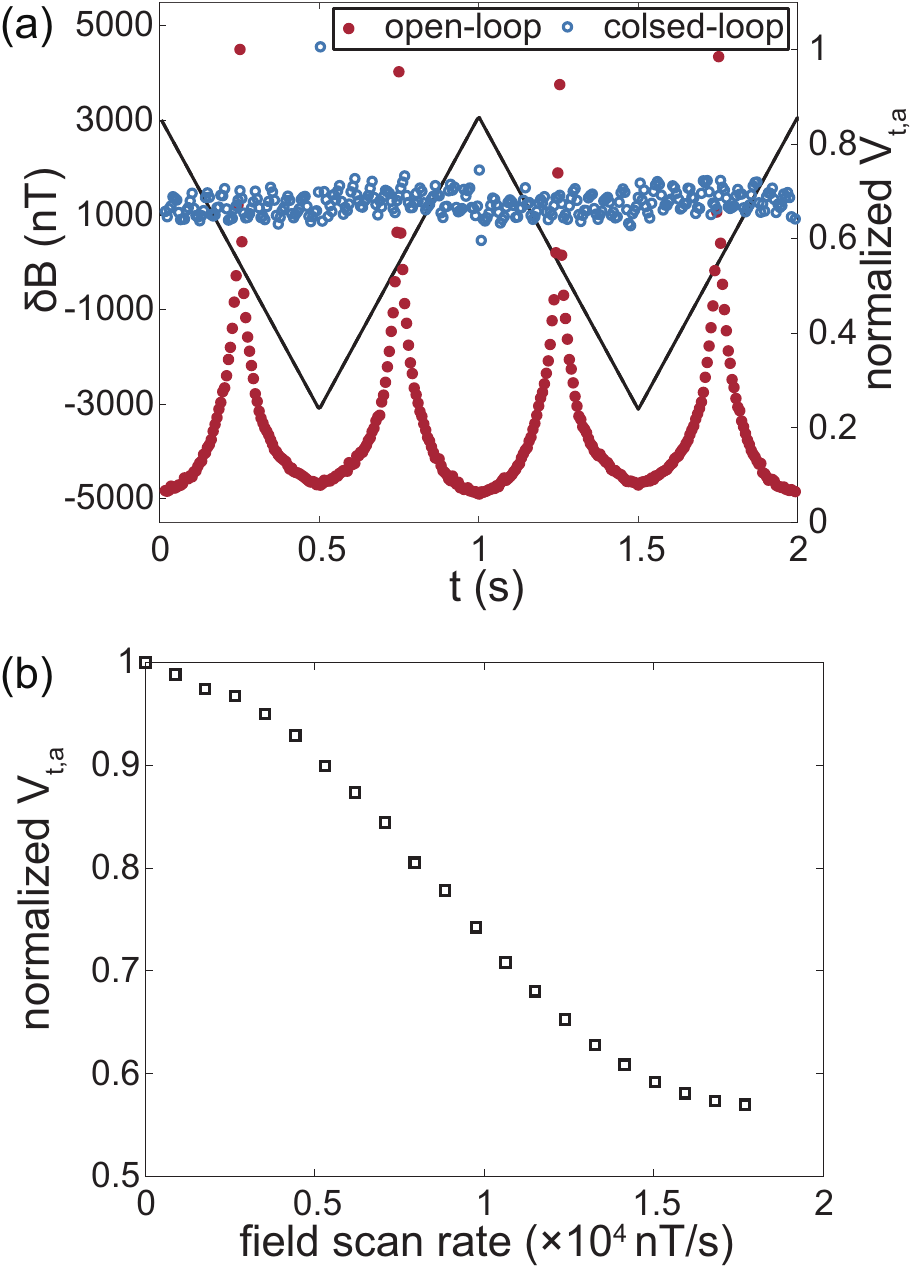}
\caption{\label{fig:sl} Plot (a) shows the ac signal amplitude ($V_{t,a}$) of the magnetometer operating in the open-loop mode (filled circle) and closed-loop mode (open circle) as a modulation field (black curve) with a frequency of 1 Hz and amplitude of 3000 nT is added on the bias field. In the closed-loop mode, the jump in $V_{t,a}$ at 0.5~s arises from the combined effects of the feedback loop delay and the reversal of the field scanning. Here, the experiment conditions are the same as these in Fig.~\ref{fig:sens}, and $V_{t,a}$ is normalized by the maximum value when the magnetometer works in the resonant conditions. Plot (b) shows the normalized $V_{t,a}$ as a function of the field scan rate for the FID magnetometer working in the closed-loop mode. }
\end{figure}

In practice, the magnitude of external field fluctuations or modulations often violates the small-amplitude assumption. Figure~\ref{fig:sl}(a) shows such an example, where the experimental conditions were the same as those in Fig.~\ref{fig:sens}(a), except that a triangle modulation with a frequency of 1 Hz and an amplitude of 3000 nT was applied on top of the bias field. Although the frequency of this modulation field is well within the magnetometer bandwidth, the modulation amplitude is so large that the ac signal amplitude is reduced by one order of magnitude during such a field scan when the magnetometer is operated in open-loop mode. The change in the ac signal amplitude is flattened when the closed-loop mode is enabled, but the resulting ac signal amplitude is still smaller than the maximum amplitude in the resonant case. This is due to the delay $\tau_d$ in the feedback loop. In the scheme presented in Fig.~\ref{fig:sig}(b), $\tau_d = T_c + T_{pu}$, where $T_c$ is the cycle time in Eq.~\eqref{eq:rhow} and $T_{pu}$ is the pump time. To further shorten the delay time, we use the first half of the signal to extract an oscillation frequency for the feedback loop, so that the Larmor frequency can be calculated before the next cycle starts. In this way, $\omega_L$ calculated from the $n$th cycle can be applied to the $\omega_d$ in the $(n+1)$th cycle, and $\tau_d$ is reduced to $T_{pu}$. The data in Fig.~\ref{fig:sl} is taken under this updated feedback strategy.

To characterize the aforementioned dynamic field response of the magnetometer, we adopt the concept of slew rate from electronics and define the field slew rate of the magnetometer as the field scan rate at which its ac signal amplitude is reduced to $1/\sqrt{2}$ of its maximum value. This parameter determines the maximum external field change rate for which {the signal-to-noise ratio of the magnetometer is above 70\% of its highest value}. As shown in Fig.~\ref{fig:sl}(b), under the given operating conditions, the slew rate of the closed-loop magnetometer is $10^4$~nT/s. To further improve this slew rate, one can reduce the pump time $T_{pu}$ or increase the pump-beam-power-broadened line width of the magnetometer response curve.

\section{Outlook}
The direct observable of an atomic magnetometer is mostly related to the atomic Larmor frequency, and this makes atomic magnetometers intrinsically scalar devices. {Taking advantage of direct frequency measurement, researchers have developed switched-bias-field schemes~\cite{alldredge1963,sapunov2006} that turn a scalar magnetometer into a vector one. The sensitivity of this vector magnetometer is given by  $(B_m/B_0)\delta B_s$, where $B_m$ is the calibrated switch-field, $B_0$ is the bias field to be measured, and $\delta B_s$ is the scalar magnetometer sensitivity.  Consequently, the high-sensitivity scalar magnetometer developed here helps reduce the requirements on $B_m$ in the vector mode. To improve the reliability of this vector magnetometer,} two important problems in the scalar one must be solved: detection dead zones and heading errors. The work presented in this paper not only presents a scheme to lift the detection dead zone but also introduces a model to explain resulting signals. The agreement between predictions of this model and experimental results for the heading error implies that such errors can be corrected to within 0.7 nT when the sensor operates in the geomagnetic field range, provided that a rough estimation of the field direction is available from the to-be-developed vector magnetometer. Thus, this work helps pave the way towards a FID vector magnetometer.

\section{Acknowledgments}
We thank helpful discussions with T. Gong and Y.-K. Feng. This work was partially carried out at the University of Science and Technology of China (USTC) Center for Micro and Nanoscale Research and Fabrication. This work is supported by National Science Foundation of China (Grant No. 12174372) and the Major Frontier Research Project of the University of Science and Technology of China (LS9990000002).

\appendix

%In this section, we present key theoretical results relevant to the present study, including unbiased frequency error estimation based on the  Cram$\mathrm{\acute{e}}$r–Rao lower bound (CRLB) and a rapid frequency estimation method utilizing the Hilbert transform.

\section{Cram$\mathrm{\acute{e}}$r–Rao lower bound (CRLB) for signals with white Gaussian noise}
Here we analyze a discretely sampled, oscillating damped signal from a single measurement, consisting of $N$ points, and the analysis follows the approach outlined in~\cite{kay1993,Gemmel2010}. For the $n$th point in the recorded data set sampled at a rate of $f_s$, the signal can be written as
\begin{equation}~\label{eq:x}
x[n]=s[n,\bm{\Theta}]+w[n]=Ae^{-n\beta}\sin(\omega n+\phi)+w[n],
\end{equation}
where $\beta=\beta_0/f_s$ with $\beta_0$ as the signal decay rate, $\omega=\omega_0/f_s$ with $\omega_0$ as the signal oscillating frequency, $s[n,\bm{\Theta}]$ represents the noise-free signal with the parameter vector $\bm{\Theta}=(A,\omega,\phi,\beta)$, and $w$ is the noise. 

The CRLB for the $i$th parameter $\theta_i$ in $\bm{\Theta}$ places a  lower bound on its variance
\begin{equation}~\label{eq:var1}
\mathrm{Var}(\theta_i)\geq [I^{-1}(\bm{\Theta})]_{ii},
\end{equation}
where $I(\bm{\Theta})$ is the Fisher information matrix. When the noise $w[n]$ is white Gaussian noise with variance $\sigma_n^2$, the elements of the Fisher information matrix are given by
\begin{equation}
I_{ij}(\bm{\Theta})=\frac{1}{\sigma_n^2}\sum_{n=0}^{N-1}\frac{\partial{s[n,\bm{\Theta}]}}{\partial \theta_i}\frac{\partial{s[n,\bm{\Theta}]}}{\partial \theta_j}.
\end{equation}

Evaluations of $I_{ij}$ involve calculating the following sums
\begin{eqnarray}
S_m&=&\sum_{n=0}^{N-1}n^me^{-2\beta n},\nonumber\\
J_m+iK_m&=&\sum_{n=0}^{N-1}n^me^{-2\beta n}e^{2i(\omega n+\phi)}.
\end{eqnarray} 
Under typical experiment conditions of $\beta\sim 10^{-3}$ and $N\beta\sim 1$, the scaling behaviors of these sums are
\begin{equation}
S_m\sim O(N^{m+1}),\quad J_m\sim O(N^m),\quad K_m\sim O(N^m). 
\end{equation}
Consequently, the scaling of the Fisher information matrix elements follows
\begin{eqnarray}
A|I_{13}|&\sim& (A/\sigma_n)^2O(1), \nonumber\\
A^2|I_{11}|, A|I_{12}|, |I_{33}|, |I_{34}|&\sim& (A/\sigma_n)^2O(N), \nonumber\\
A|I_{14}|, |I_{23}|, |I_{24}|&\sim& (A/\sigma_n)^2O(N^2), \nonumber\\
|I_{22}|, |I_{44}| &\sim&(A/\sigma_n)^2O(N^3).
\end{eqnarray}
Keeping only the dominant terms, we get the inverse Fisher information element corresponding to the parameter of oscillation frequency as
\begin{equation}~\label{eq:I-1}
[I^{-1}(\bm{\Theta})]_{22}=\frac{I_{33}}{I_{22}I_{33}-I_{23}^2}=\frac{2}{(A/\sigma)^2}\frac{S_0}{S_0S_2-S_1^2}.
\end{equation}
The combination of Eqs.~\eqref{eq:var1} and~\eqref{eq:I-1} yields the CRLB for the oscillation frequency from a single measurement as
\begin{equation}~\label{eq:vars}
\mathrm{Var}_s(\omega_0)=\frac{2f_s^2}{(A/\sigma_n)^2}\frac{(1-z^2)^3(1-z^{2N})}{z^2(1-z^{2N})^2-N^2z^{2N}(1-z^2)^2},
\end{equation}
with $z=e^{-\beta}$. 

For white noise, its time-domain variance $\sigma_n^2$ is connected with noise power spectral density $\rho_n^2$ by $\sigma_n^2=f_s\rho_n^2/2$. Substituting this relation gives an alternative form of the CRLB of the frequency parameter for a single measurement:
\begin{equation}
\mathrm{Var}_s(\omega_0)=\frac{f_s^3}{(A/\rho_n)^2}\frac{(1-z^2)^3(1-z^{2N})}{z^2(1-z^{2N})^2-N^2z^{2N}(1-z^2)^2}.
\end{equation}
Finally, if the measurement is repeated with cycle time $T_c$,  the sensitivity for estimating $\omega_0$, detnoted as $\rho_{\omega}$, becomes
\begin{equation}~\label{eq:rhow}
\rho_{\omega}=\sqrt{2\mathrm{Var}_s(\omega_0)T_c}.
\end{equation}

\section{Frequency calculation based on Hilbert transform}
The most reliable approach for extracting oscillation frequencies from FID signals involves nonlinear fitting with established algorithms, such as the Levenberg–Marquardt method~\cite{press2007}. However, this method requires intense computation time and prior knowledge of the signal’s analytic expression. In addition, its results can be sensitive to the choice of initial values of parameters. As an alternative, we have recently developed a robust and rapid technique based on the Hilbert transform that achieves precision comparable to nonlinear fitting method for FID signals~\cite{gong2025}. In the following, we briefly outline the key concepts of this new approach.

For the signal in Eq.~\eqref{eq:x}, a 90$^\circ$ phase shifter can be applied to generate a new sequence 
\begin{equation}~\label{eq:y}
y[n]=\sum_{m=-\infty}^\infty h[m]x[n-m].
\end{equation}
Here, $h[m]$ is the impulse response of the 90$^\circ$ phase shifter, given by~\cite{oppenheim2010}
\begin{equation}
h[m]=
\begin{cases}
\frac{2}{\pi}\frac{\sin^2(m\pi/2)}{m}, & m\neq0, \\
0, & m=0.
\end{cases}
\end{equation}
In practical implementations, the summation is truncated to $|m|=15$, a choice that balances computational speed with sufficient precision. 

From the data sequences $x[n]$ and $y[n]$, the instantaneous phase of the oscillating component in $x[n]$ is obtained as
\begin{equation} 
\varphi[n]=\arctan\left(x[n]/y[n]\right)+k\pi, 
\end{equation}
where integer multiples of $\pi$ are added to maintain phase continuity. The oscillation frequency $\omega_0$ is then extracted by performing a linear fit of $\varphi[n]$ as a function of time.


\begin{thebibliography}{20}%
\makeatletter
\providecommand \@ifxundefined [1]{%
 \@ifx{#1\undefined}
}%
\providecommand \@ifnum [1]{%
 \ifnum #1\expandafter \@firstoftwo
 \else \expandafter \@secondoftwo
 \fi
}%
\providecommand \@ifx [1]{%
 \ifx #1\expandafter \@firstoftwo
 \else \expandafter \@secondoftwo
 \fi
}%
\providecommand \natexlab [1]{#1}%
\providecommand \enquote  [1]{``#1''}%
\providecommand \bibnamefont  [1]{#1}%
\providecommand \bibfnamefont [1]{#1}%
\providecommand \citenamefont [1]{#1}%
\providecommand \href@noop [0]{\@secondoftwo}%
\providecommand \href [0]{\begingroup \@sanitize@url \@href}%
\providecommand \@href[1]{\@@startlink{#1}\@@href}%
\providecommand \@@href[1]{\endgroup#1\@@endlink}%
\providecommand \@sanitize@url [0]{\catcode `\\12\catcode `\$12\catcode
  `\&12\catcode `\#12\catcode `\^12\catcode `\_12\catcode `\%12\relax}%
\providecommand \@@startlink[1]{}%
\providecommand \@@endlink[0]{}%
\providecommand \url  [0]{\begingroup\@sanitize@url \@url }%
\providecommand \@url [1]{\endgroup\@href {#1}{\urlprefix }}%
\providecommand \urlprefix  [0]{URL }%
\providecommand \Eprint [0]{\href }%
\providecommand \doibase [0]{https://doi.org/}%
\providecommand \selectlanguage [0]{\@gobble}%
\providecommand \bibinfo  [0]{\@secondoftwo}%
\providecommand \bibfield  [0]{\@secondoftwo}%
\providecommand \translation [1]{[#1]}%
\providecommand \BibitemOpen [0]{}%
\providecommand \bibitemStop [0]{}%
\providecommand \bibitemNoStop [0]{.\EOS\space}%
\providecommand \EOS [0]{\spacefactor3000\relax}%
\providecommand \BibitemShut  [1]{\csname bibitem#1\endcsname}%
\let\auto@bib@innerbib\@empty
%</preamble>
\bibitem [{\citenamefont {Jackson~Kimball}\ \emph {et~al.}(2017)\citenamefont
  {Jackson~Kimball}, \citenamefont {Dudley}, \citenamefont {Li}, \citenamefont
  {Patel},\ and\ \citenamefont {Valdez}}]{kimball2017}%
  \BibitemOpen
  \bibfield  {author} {\bibinfo {author} {\bibfnamefont {D.~F.}\ \bibnamefont
  {Jackson~Kimball}}, \bibinfo {author} {\bibfnamefont {J.}~\bibnamefont
  {Dudley}}, \bibinfo {author} {\bibfnamefont {Y.}~\bibnamefont {Li}}, \bibinfo
  {author} {\bibfnamefont {D.}~\bibnamefont {Patel}},\ and\ \bibinfo {author}
  {\bibfnamefont {J.}~\bibnamefont {Valdez}},\ }\bibfield  {title} {\bibinfo
  {title} {Constraints on long-range spin-gravity and monopole-dipole couplings
  of the proton},\ }\href {https://doi.org/10.1103/PhysRevD.96.075004}
  {\bibfield  {journal} {\bibinfo  {journal} {Phys. Rev. D}\ }\textbf {\bibinfo
  {volume} {96}},\ \bibinfo {pages} {075004} (\bibinfo {year}
  {2017})}\BibitemShut {NoStop}%
\bibitem [{\citenamefont {Abel}\ \emph {et~al.}(2020)\citenamefont {Abel},
  \citenamefont {Afach}, \citenamefont {Ayres}, \citenamefont {Ban},
  \citenamefont {Bison}, \citenamefont {Bodek}, \citenamefont {Bondar},
  \citenamefont {Chanel}, \citenamefont {Chiu}, \citenamefont {Crawford},
  \citenamefont {Chowdhuri}, \citenamefont {Daum}, \citenamefont {Emmenegger},
  \citenamefont {Ferraris-Bouchez}, \citenamefont {Fertl}, \citenamefont
  {Franke}, \citenamefont {Griffith}, \citenamefont
  {Gruji\ifmmode~\acute{c}\else \'{c}\fi{}}, \citenamefont {Hayen},
  \citenamefont {H\'elaine}, \citenamefont {Hild}, \citenamefont {Kasprzak},
  \citenamefont {Kermaidic}, \citenamefont {Kirch}, \citenamefont {Knowles},
  \citenamefont {Koch}, \citenamefont {Komposch}, \citenamefont {Koss},
  \citenamefont {Kozela}, \citenamefont {Krempel}, \citenamefont {Lauss},
  \citenamefont {Lefort}, \citenamefont {Lemi\`ere}, \citenamefont {Leredde},
  \citenamefont {Mtchedlishvili}, \citenamefont {Mohanmurthy}, \citenamefont
  {Musgrave}, \citenamefont {Naviliat-Cuncic}, \citenamefont {Pais},
  \citenamefont {Pazgalev}, \citenamefont {Piegsa}, \citenamefont {Pierre},
  \citenamefont {Pignol}, \citenamefont {Prashanth}, \citenamefont
  {Qu\'em\'ener}, \citenamefont {Rawlik}, \citenamefont {Rebreyend},
  \citenamefont {Ries}, \citenamefont {Roccia}, \citenamefont {Rozpedzik},
  \citenamefont {Schmidt-Wellenburg}, \citenamefont {Schnabel}, \citenamefont
  {Severijns}, \citenamefont {Dinani}, \citenamefont {Thorne}, \citenamefont
  {Weis}, \citenamefont {Wursten}, \citenamefont {Wyszynski}, \citenamefont
  {Zejma},\ and\ \citenamefont {Zsigmond}}]{abel2020}%
  \BibitemOpen
  \bibfield  {author} {\bibinfo {author} {\bibfnamefont {C.}~\bibnamefont
  {Abel}}, \bibinfo {author} {\bibfnamefont {S.}~\bibnamefont {Afach}},
  \bibinfo {author} {\bibfnamefont {N.~J.}\ \bibnamefont {Ayres}}, \bibinfo
  {author} {\bibfnamefont {G.}~\bibnamefont {Ban}}, \bibinfo {author}
  {\bibfnamefont {G.}~\bibnamefont {Bison}}, \bibinfo {author} {\bibfnamefont
  {K.}~\bibnamefont {Bodek}}, \bibinfo {author} {\bibfnamefont
  {V.}~\bibnamefont {Bondar}}, \bibinfo {author} {\bibfnamefont
  {E.}~\bibnamefont {Chanel}}, \bibinfo {author} {\bibfnamefont {P.-J.}\
  \bibnamefont {Chiu}}, \bibinfo {author} {\bibfnamefont {C.~B.}\ \bibnamefont
  {Crawford}}, \bibinfo {author} {\bibfnamefont {Z.}~\bibnamefont {Chowdhuri}},
  \bibinfo {author} {\bibfnamefont {M.}~\bibnamefont {Daum}}, \bibinfo {author}
  {\bibfnamefont {S.}~\bibnamefont {Emmenegger}}, \bibinfo {author}
  {\bibfnamefont {L.}~\bibnamefont {Ferraris-Bouchez}}, \bibinfo {author}
  {\bibfnamefont {M.}~\bibnamefont {Fertl}}, \bibinfo {author} {\bibfnamefont
  {B.}~\bibnamefont {Franke}}, \bibinfo {author} {\bibfnamefont {W.~C.}\
  \bibnamefont {Griffith}}, \bibinfo {author} {\bibfnamefont {Z.~D.}\
  \bibnamefont {Gruji\ifmmode~\acute{c}\else \'{c}\fi{}}}, \bibinfo {author}
  {\bibfnamefont {L.}~\bibnamefont {Hayen}}, \bibinfo {author} {\bibfnamefont
  {V.}~\bibnamefont {H\'elaine}}, \bibinfo {author} {\bibfnamefont
  {N.}~\bibnamefont {Hild}}, \bibinfo {author} {\bibfnamefont {M.}~\bibnamefont
  {Kasprzak}}, \bibinfo {author} {\bibfnamefont {Y.}~\bibnamefont {Kermaidic}},
  \bibinfo {author} {\bibfnamefont {K.}~\bibnamefont {Kirch}}, \bibinfo
  {author} {\bibfnamefont {P.}~\bibnamefont {Knowles}}, \bibinfo {author}
  {\bibfnamefont {H.-C.}\ \bibnamefont {Koch}}, \bibinfo {author}
  {\bibfnamefont {S.}~\bibnamefont {Komposch}}, \bibinfo {author}
  {\bibfnamefont {P.~A.}\ \bibnamefont {Koss}}, \bibinfo {author}
  {\bibfnamefont {A.}~\bibnamefont {Kozela}}, \bibinfo {author} {\bibfnamefont
  {J.}~\bibnamefont {Krempel}}, \bibinfo {author} {\bibfnamefont
  {B.}~\bibnamefont {Lauss}}, \bibinfo {author} {\bibfnamefont
  {T.}~\bibnamefont {Lefort}}, \bibinfo {author} {\bibfnamefont
  {Y.}~\bibnamefont {Lemi\`ere}}, \bibinfo {author} {\bibfnamefont
  {A.}~\bibnamefont {Leredde}}, \bibinfo {author} {\bibfnamefont
  {A.}~\bibnamefont {Mtchedlishvili}}, \bibinfo {author} {\bibfnamefont
  {P.}~\bibnamefont {Mohanmurthy}}, \bibinfo {author} {\bibfnamefont
  {M.}~\bibnamefont {Musgrave}}, \bibinfo {author} {\bibfnamefont
  {O.}~\bibnamefont {Naviliat-Cuncic}}, \bibinfo {author} {\bibfnamefont
  {D.}~\bibnamefont {Pais}}, \bibinfo {author} {\bibfnamefont {A.}~\bibnamefont
  {Pazgalev}}, \bibinfo {author} {\bibfnamefont {F.~M.}\ \bibnamefont
  {Piegsa}}, \bibinfo {author} {\bibfnamefont {E.}~\bibnamefont {Pierre}},
  \bibinfo {author} {\bibfnamefont {G.}~\bibnamefont {Pignol}}, \bibinfo
  {author} {\bibfnamefont {P.~N.}\ \bibnamefont {Prashanth}}, \bibinfo {author}
  {\bibfnamefont {G.}~\bibnamefont {Qu\'em\'ener}}, \bibinfo {author}
  {\bibfnamefont {M.}~\bibnamefont {Rawlik}}, \bibinfo {author} {\bibfnamefont
  {D.}~\bibnamefont {Rebreyend}}, \bibinfo {author} {\bibfnamefont
  {D.}~\bibnamefont {Ries}}, \bibinfo {author} {\bibfnamefont {S.}~\bibnamefont
  {Roccia}}, \bibinfo {author} {\bibfnamefont {D.}~\bibnamefont {Rozpedzik}},
  \bibinfo {author} {\bibfnamefont {P.}~\bibnamefont {Schmidt-Wellenburg}},
  \bibinfo {author} {\bibfnamefont {A.}~\bibnamefont {Schnabel}}, \bibinfo
  {author} {\bibfnamefont {N.}~\bibnamefont {Severijns}}, \bibinfo {author}
  {\bibfnamefont {R.~T.}\ \bibnamefont {Dinani}}, \bibinfo {author}
  {\bibfnamefont {J.}~\bibnamefont {Thorne}}, \bibinfo {author} {\bibfnamefont
  {A.}~\bibnamefont {Weis}}, \bibinfo {author} {\bibfnamefont {E.}~\bibnamefont
  {Wursten}}, \bibinfo {author} {\bibfnamefont {G.}~\bibnamefont {Wyszynski}},
  \bibinfo {author} {\bibfnamefont {J.}~\bibnamefont {Zejma}},\ and\ \bibinfo
  {author} {\bibfnamefont {G.}~\bibnamefont {Zsigmond}},\ }\bibfield  {title}
  {\bibinfo {title} {Optically pumped {C}s magnetometers enabling a
  high-sensitivity search for the neutron electric dipole moment},\ }\href
  {https://doi.org/10.1103/PhysRevA.101.053419} {\bibfield  {journal} {\bibinfo
   {journal} {Phys. Rev. A}\ }\textbf {\bibinfo {volume} {101}},\ \bibinfo
  {pages} {053419} (\bibinfo {year} {2020})}\BibitemShut {NoStop}%
\bibitem [{\citenamefont {Nightingale}\ \emph {et~al.}(2025)\citenamefont
  {Nightingale}, \citenamefont {Coussens}, \citenamefont {Woodley},
  \citenamefont {Nicolau}, \citenamefont {Page}, \citenamefont {Kruger},\ and\
  \citenamefont {Orucevic}}]{nightingale2025}%
  \BibitemOpen
  \bibfield  {author} {\bibinfo {author} {\bibfnamefont {D.}~\bibnamefont
  {Nightingale}}, \bibinfo {author} {\bibfnamefont {T.}~\bibnamefont
  {Coussens}}, \bibinfo {author} {\bibfnamefont {M.~T.~M.}\ \bibnamefont
  {Woodley}}, \bibinfo {author} {\bibfnamefont {D.}~\bibnamefont {Nicolau}},
  \bibinfo {author} {\bibfnamefont {L.}~\bibnamefont {Page}}, \bibinfo {author}
  {\bibfnamefont {P.}~\bibnamefont {Kruger}},\ and\ \bibinfo {author}
  {\bibfnamefont {F.}~\bibnamefont {Orucevic}},\ }\bibfield  {title} {\bibinfo
  {title} {Mobile total field optically pumped magnetometers for navigation},\
  }in\ \href@noop {} {\emph {\bibinfo {booktitle} {Workshop on Optically Pumped
  Magnetometers}}}\ (\bibinfo {year} {2025})\BibitemShut {NoStop}%
\bibitem [{\citenamefont {Limes}\ \emph {et~al.}(2020)\citenamefont {Limes},
  \citenamefont {Foley}, \citenamefont {Kornack}, \citenamefont {Caliga},
  \citenamefont {McBride}, \citenamefont {Braun}, \citenamefont {Lee},
  \citenamefont {Lucivero},\ and\ \citenamefont {Romalis}}]{limes2020}%
  \BibitemOpen
  \bibfield  {author} {\bibinfo {author} {\bibfnamefont {M.}~\bibnamefont
  {Limes}}, \bibinfo {author} {\bibfnamefont {E.}~\bibnamefont {Foley}},
  \bibinfo {author} {\bibfnamefont {T.}~\bibnamefont {Kornack}}, \bibinfo
  {author} {\bibfnamefont {S.}~\bibnamefont {Caliga}}, \bibinfo {author}
  {\bibfnamefont {S.}~\bibnamefont {McBride}}, \bibinfo {author} {\bibfnamefont
  {A.}~\bibnamefont {Braun}}, \bibinfo {author} {\bibfnamefont
  {W.}~\bibnamefont {Lee}}, \bibinfo {author} {\bibfnamefont {V.}~\bibnamefont
  {Lucivero}},\ and\ \bibinfo {author} {\bibfnamefont {M.}~\bibnamefont
  {Romalis}},\ }\bibfield  {title} {\bibinfo {title} {Portable magnetometry for
  detection of biomagnetism in ambient environments},\ }\href
  {https://doi.org/10.1103/PhysRevApplied.14.011002} {\bibfield  {journal}
  {\bibinfo  {journal} {Phys. Rev. Appl.}\ }\textbf {\bibinfo {volume} {14}},\
  \bibinfo {pages} {011002} (\bibinfo {year} {2020})}\BibitemShut {NoStop}%
\bibitem [{\citenamefont {Lee}\ \emph {et~al.}(2021)\citenamefont {Lee},
  \citenamefont {Lucivero}, \citenamefont {Romalis}, \citenamefont {Limes},
  \citenamefont {Foley},\ and\ \citenamefont {Kornack}}]{lee2021}%
  \BibitemOpen
  \bibfield  {author} {\bibinfo {author} {\bibfnamefont {W.}~\bibnamefont
  {Lee}}, \bibinfo {author} {\bibfnamefont {V.~G.}\ \bibnamefont {Lucivero}},
  \bibinfo {author} {\bibfnamefont {M.~V.}\ \bibnamefont {Romalis}}, \bibinfo
  {author} {\bibfnamefont {M.~E.}\ \bibnamefont {Limes}}, \bibinfo {author}
  {\bibfnamefont {E.~L.}\ \bibnamefont {Foley}},\ and\ \bibinfo {author}
  {\bibfnamefont {T.~W.}\ \bibnamefont {Kornack}},\ }\bibfield  {title}
  {\bibinfo {title} {Heading errors in all-optical alkali-metal-vapor
  magnetometers in geomagnetic fields},\ }\href
  {https://doi.org/10.1103/PhysRevA.103.063103} {\bibfield  {journal} {\bibinfo
   {journal} {Phys. Rev. A}\ }\textbf {\bibinfo {volume} {103}},\ \bibinfo
  {pages} {063103} (\bibinfo {year} {2021})}\BibitemShut {NoStop}%
\bibitem [{\citenamefont {Liu}\ \emph {et~al.}(2025)\citenamefont {Liu},
  \citenamefont {Wang}, \citenamefont {Zhang}, \citenamefont {Xiao},\ and\
  \citenamefont {Sheng}}]{liu2025}%
  \BibitemOpen
  \bibfield  {author} {\bibinfo {author} {\bibfnamefont {S.-Q.}\ \bibnamefont
  {Liu}}, \bibinfo {author} {\bibfnamefont {X.-K.}\ \bibnamefont {Wang}},
  \bibinfo {author} {\bibfnamefont {X.-D.}\ \bibnamefont {Zhang}}, \bibinfo
  {author} {\bibfnamefont {W.}~\bibnamefont {Xiao}},\ and\ \bibinfo {author}
  {\bibfnamefont {D.}~\bibnamefont {Sheng}},\ }\bibfield  {title} {\bibinfo
  {title} {Suppression of heading errors in bell-bloom optically pumped
  free-induction-decay alkali-metal atomic magnetometers},\ }\href
  {https://doi.org/10.1103/PhysRevA.111.023119} {\bibfield  {journal} {\bibinfo
   {journal} {Phys. Rev. A}\ }\textbf {\bibinfo {volume} {111}},\ \bibinfo
  {pages} {023119} (\bibinfo {year} {2025})}\BibitemShut {NoStop}%
\bibitem [{\citenamefont {Gong}\ \emph {et~al.}(2025)\citenamefont {Gong},
  \citenamefont {Shu}, \citenamefont {He}, \citenamefont {Liu}, \citenamefont
  {Li}, \citenamefont {Hao}, \citenamefont {Sheng}, \citenamefont {Wang},\ and\
  \citenamefont {Feng}}]{gong2025}%
  \BibitemOpen
  \bibfield  {author} {\bibinfo {author} {\bibfnamefont {T.}~\bibnamefont
  {Gong}}, \bibinfo {author} {\bibfnamefont {M.-R.}\ \bibnamefont {Shu}},
  \bibinfo {author} {\bibfnamefont {J.}~\bibnamefont {He}}, \bibinfo {author}
  {\bibfnamefont {K.}~\bibnamefont {Liu}}, \bibinfo {author} {\bibfnamefont
  {Y.-R.}\ \bibnamefont {Li}}, \bibinfo {author} {\bibfnamefont {X.-J.}\
  \bibnamefont {Hao}}, \bibinfo {author} {\bibfnamefont {D.}~\bibnamefont
  {Sheng}}, \bibinfo {author} {\bibfnamefont {Y.-M.}\ \bibnamefont {Wang}},\
  and\ \bibinfo {author} {\bibfnamefont {Y.-K.}\ \bibnamefont {Feng}},\
  }\bibfield  {title} {\bibinfo {title} {A high-sensitivity frequency counter
  for free-induction-decay signals},\ }\href
  {https://doi.org/10.1109/TIM.2025.3575977} {\bibfield  {journal} {\bibinfo
  {journal} {IEEE Transactions on Instrumentation and Measurement}\ }\textbf
  {\bibinfo {volume} {74}},\ \bibinfo {pages} {1} (\bibinfo {year}
  {2025})}\BibitemShut {NoStop}%
\bibitem [{\citenamefont {Ben-Kish}\ and\ \citenamefont
  {Romalis}(2010)}]{benkish2010}%
  \BibitemOpen
  \bibfield  {author} {\bibinfo {author} {\bibfnamefont {A.}~\bibnamefont
  {Ben-Kish}}\ and\ \bibinfo {author} {\bibfnamefont {M.~V.}\ \bibnamefont
  {Romalis}},\ }\bibfield  {title} {\bibinfo {title} {Dead-zone-free atomic
  magnetometry with simultaneous excitation of orientation and alignment
  resonances},\ }\href {https://doi.org/10.1103/PhysRevLett.105.193601}
  {\bibfield  {journal} {\bibinfo  {journal} {Phys. Rev. Lett.}\ }\textbf
  {\bibinfo {volume} {105}},\ \bibinfo {pages} {193601} (\bibinfo {year}
  {2010})}\BibitemShut {NoStop}%
\bibitem [{\citenamefont {Mehta}\ \emph {et~al.}(2025)\citenamefont {Mehta},
  \citenamefont {Samanta},\ and\ \citenamefont {Grewal}}]{mehta2025}%
  \BibitemOpen
  \bibfield  {author} {\bibinfo {author} {\bibfnamefont {S.}~\bibnamefont
  {Mehta}}, \bibinfo {author} {\bibfnamefont {G.~K.}\ \bibnamefont {Samanta}},\
  and\ \bibinfo {author} {\bibfnamefont {R.~S.}\ \bibnamefont {Grewal}},\
  }\bibfield  {title} {\bibinfo {title} {Dead-zone-free single-beam atomic
  magnetometer based on free-induction-decay of rb atoms},\ }\href
  {https://doi.org/10.1063/5.0248330} {\bibfield  {journal} {\bibinfo
  {journal} {Applied Physics Letters}\ }\textbf {\bibinfo {volume} {126}},\
  \bibinfo {pages} {044002} (\bibinfo {year} {2025})}\BibitemShut {NoStop}%
\bibitem [{\citenamefont {Ness}(1970)}]{ness1970}%
  \BibitemOpen
  \bibfield  {author} {\bibinfo {author} {\bibfnamefont {N.~F.}\ \bibnamefont
  {Ness}},\ }\bibfield  {title} {\bibinfo {title} {Magnetometers for space
  research},\ }\href {https://doi.org/10.1007/BF00183028} {\bibfield  {journal}
  {\bibinfo  {journal} {Space Science Reviews}\ }\textbf {\bibinfo {volume}
  {11}},\ \bibinfo {pages} {459} (\bibinfo {year} {1970})}\BibitemShut
  {NoStop}%
\bibitem [{\citenamefont {Yu}\ \emph {et~al.}(2022)\citenamefont {Yu},
  \citenamefont {Liu}, \citenamefont {Yuan},\ and\ \citenamefont
  {Sheng}}]{yu2022}%
  \BibitemOpen
  \bibfield  {author} {\bibinfo {author} {\bibfnamefont {Q.-Q.}\ \bibnamefont
  {Yu}}, \bibinfo {author} {\bibfnamefont {S.-Q.}\ \bibnamefont {Liu}},
  \bibinfo {author} {\bibfnamefont {C.-Q.}\ \bibnamefont {Yuan}},\ and\
  \bibinfo {author} {\bibfnamefont {D.}~\bibnamefont {Sheng}},\ }\bibfield
  {title} {\bibinfo {title} {Light-shift-free and dead-zone-free
  atomic-orientation-based scalar magnetometry using a single
  amplitude-modulated beam},\ }\href
  {https://doi.org/10.1103/PhysRevApplied.18.014015} {\bibfield  {journal}
  {\bibinfo  {journal} {Phys. Rev. Appl.}\ }\textbf {\bibinfo {volume} {18}},\
  \bibinfo {pages} {014015} (\bibinfo {year} {2022})}\BibitemShut {NoStop}%
\bibitem [{\citenamefont {Bell}\ and\ \citenamefont {Bloom}(1961)}]{bell1961}%
  \BibitemOpen
  \bibfield  {author} {\bibinfo {author} {\bibfnamefont {W.~E.}\ \bibnamefont
  {Bell}}\ and\ \bibinfo {author} {\bibfnamefont {A.~L.}\ \bibnamefont
  {Bloom}},\ }\bibfield  {title} {\bibinfo {title} {Optically driven spin
  precession},\ }\href {https://doi.org/10.1103/PhysRevLett.6.280} {\bibfield
  {journal} {\bibinfo  {journal} {Phys. Rev. Lett.}\ }\textbf {\bibinfo
  {volume} {6}},\ \bibinfo {pages} {280} (\bibinfo {year} {1961})}\BibitemShut
  {NoStop}%
\bibitem [{\citenamefont {Appelt}\ \emph {et~al.}(1998)\citenamefont {Appelt},
  \citenamefont {Baranga}, \citenamefont {Erickson}, \citenamefont {Romalis},
  \citenamefont {Young},\ and\ \citenamefont {Happer}}]{appelt98}%
  \BibitemOpen
  \bibfield  {author} {\bibinfo {author} {\bibfnamefont {S.}~\bibnamefont
  {Appelt}}, \bibinfo {author} {\bibfnamefont {A.~B.-A.}\ \bibnamefont
  {Baranga}}, \bibinfo {author} {\bibfnamefont {C.~J.}\ \bibnamefont
  {Erickson}}, \bibinfo {author} {\bibfnamefont {M.~V.}\ \bibnamefont
  {Romalis}}, \bibinfo {author} {\bibfnamefont {A.~R.}\ \bibnamefont {Young}},\
  and\ \bibinfo {author} {\bibfnamefont {W.}~\bibnamefont {Happer}},\
  }\bibfield  {title} {\bibinfo {title} {Theory of spin-exchange optical
  pumping of ${}^{3}\mathrm{He}$ and ${}^{129}\mathrm{Xe}$},\ }\href
  {https://doi.org/10.1103/PhysRevA.58.1412} {\bibfield  {journal} {\bibinfo
  {journal} {Phys. Rev. A}\ }\textbf {\bibinfo {volume} {58}},\ \bibinfo
  {pages} {1412} (\bibinfo {year} {1998})}\BibitemShut {NoStop}%
\bibitem [{\citenamefont {Yi}\ \emph {et~al.}(2024)\citenamefont {Yi},
  \citenamefont {Liu}, \citenamefont {Wang}, \citenamefont {Xiao},
  \citenamefont {Sheng}, \citenamefont {Peng},\ and\ \citenamefont
  {Guo}}]{yi2024}%
  \BibitemOpen
  \bibfield  {author} {\bibinfo {author} {\bibfnamefont {K.}~\bibnamefont
  {Yi}}, \bibinfo {author} {\bibfnamefont {Y.}~\bibnamefont {Liu}}, \bibinfo
  {author} {\bibfnamefont {B.}~\bibnamefont {Wang}}, \bibinfo {author}
  {\bibfnamefont {W.}~\bibnamefont {Xiao}}, \bibinfo {author} {\bibfnamefont
  {D.}~\bibnamefont {Sheng}}, \bibinfo {author} {\bibfnamefont
  {X.}~\bibnamefont {Peng}},\ and\ \bibinfo {author} {\bibfnamefont
  {H.}~\bibnamefont {Guo}},\ }\bibfield  {title} {\bibinfo {title}
  {Free-induction-decay ${}^{4}\mathrm{He}$ magnetometer using a multipass
  cell},\ }\href {https://doi.org/10.1103/PhysRevApplied.22.014084} {\bibfield
  {journal} {\bibinfo  {journal} {Phys. Rev. Appl.}\ }\textbf {\bibinfo
  {volume} {22}},\ \bibinfo {pages} {014084} (\bibinfo {year}
  {2024})}\BibitemShut {NoStop}%
\bibitem [{\citenamefont {Alldredge}\ and\ \citenamefont
  {Saldukas}(1964)}]{alldredge1963}%
  \BibitemOpen
  \bibfield  {author} {\bibinfo {author} {\bibfnamefont {L.~R.}\ \bibnamefont
  {Alldredge}}\ and\ \bibinfo {author} {\bibfnamefont {I.}~\bibnamefont
  {Saldukas}},\ }\bibfield  {title} {\bibinfo {title} {An automatic standard
  magnetic observatory},\ }\href
  {https://doi.org/https://doi.org/10.1029/JZ069i010p01963} {\bibfield
  {journal} {\bibinfo  {journal} {Journal of Geophysical Research (1896-1977)}\
  }\textbf {\bibinfo {volume} {69}},\ \bibinfo {pages} {1963} (\bibinfo {year}
  {1964})}\BibitemShut {NoStop}%
\bibitem [{\citenamefont {Sapunov}\ \emph {et~al.}(2006)\citenamefont
  {Sapunov}, \citenamefont {Rasson}, \citenamefont {Denisov}, \citenamefont
  {Saveliev}, \citenamefont {Kiselev}, \citenamefont {Denisova}, \citenamefont
  {Podmogov},\ and\ \citenamefont {Khomutov}}]{sapunov2006}%
  \BibitemOpen
  \bibfield  {author} {\bibinfo {author} {\bibfnamefont {V.}~\bibnamefont
  {Sapunov}}, \bibinfo {author} {\bibfnamefont {J.}~\bibnamefont {Rasson}},
  \bibinfo {author} {\bibfnamefont {A.}~\bibnamefont {Denisov}}, \bibinfo
  {author} {\bibfnamefont {D.}~\bibnamefont {Saveliev}}, \bibinfo {author}
  {\bibfnamefont {S.}~\bibnamefont {Kiselev}}, \bibinfo {author} {\bibfnamefont
  {O.}~\bibnamefont {Denisova}}, \bibinfo {author} {\bibfnamefont
  {Y.}~\bibnamefont {Podmogov}},\ and\ \bibinfo {author} {\bibfnamefont
  {S.}~\bibnamefont {Khomutov}},\ }\bibfield  {title} {\bibinfo {title}
  {Theodolite-borne vector overhauser magnetometer: Dimover},\ }\href
  {https://doi.org/10.1186/BF03351972} {\bibfield  {journal} {\bibinfo
  {journal} {Earth, Planets and Space}\ }\textbf {\bibinfo {volume} {58}},\
  \bibinfo {pages} {711} (\bibinfo {year} {2006})}\BibitemShut {NoStop}%
\bibitem [{\citenamefont {Kay}(1993)}]{kay1993}%
  \BibitemOpen
  \bibfield  {author} {\bibinfo {author} {\bibfnamefont {S.~M.}\ \bibnamefont
  {Kay}},\ }\href@noop {} {\emph {\bibinfo {title} {Fundamentals of Statistical
  Signal Processing: Estimation Theory}}},\ Vol.~\bibinfo {volume} {I}\
  (\bibinfo  {publisher} {Pearson},\ \bibinfo {year} {1993})\BibitemShut
  {NoStop}%
\bibitem [{\citenamefont {Gemmel}\ \emph {et~al.}(2010)\citenamefont {Gemmel},
  \citenamefont {Heil}, \citenamefont {Karpuk}, \citenamefont {Lenz},
  \citenamefont {Ludwig}, \citenamefont {Sobolev}, \citenamefont {Tullney},
  \citenamefont {Burghoff}, \citenamefont {Kilian}, \citenamefont
  {Knappe-Grüneberg}, \citenamefont {Müller}, \citenamefont {Schnabel},
  \citenamefont {Seifert}, \citenamefont {Trahms},\ and\ \citenamefont
  {Baeßler}}]{Gemmel2010}%
  \BibitemOpen
  \bibfield  {author} {\bibinfo {author} {\bibfnamefont {C.}~\bibnamefont
  {Gemmel}}, \bibinfo {author} {\bibfnamefont {W.}~\bibnamefont {Heil}},
  \bibinfo {author} {\bibfnamefont {S.}~\bibnamefont {Karpuk}}, \bibinfo
  {author} {\bibfnamefont {K.}~\bibnamefont {Lenz}}, \bibinfo {author}
  {\bibfnamefont {C.}~\bibnamefont {Ludwig}}, \bibinfo {author} {\bibfnamefont
  {Y.}~\bibnamefont {Sobolev}}, \bibinfo {author} {\bibfnamefont
  {K.}~\bibnamefont {Tullney}}, \bibinfo {author} {\bibfnamefont
  {M.}~\bibnamefont {Burghoff}}, \bibinfo {author} {\bibfnamefont
  {W.}~\bibnamefont {Kilian}}, \bibinfo {author} {\bibfnamefont
  {S.}~\bibnamefont {Knappe-Grüneberg}}, \bibinfo {author} {\bibfnamefont
  {W.}~\bibnamefont {Müller}}, \bibinfo {author} {\bibfnamefont
  {A.}~\bibnamefont {Schnabel}}, \bibinfo {author} {\bibfnamefont
  {F.}~\bibnamefont {Seifert}}, \bibinfo {author} {\bibfnamefont
  {L.}~\bibnamefont {Trahms}},\ and\ \bibinfo {author} {\bibfnamefont
  {S.}~\bibnamefont {Baeßler}},\ }\bibfield  {title} {\bibinfo {title}
  {Ultra-sensitive magnetometry based on free precession of nuclear spins},\
  }\href {https://doi.org/10.1140/epjd/e2010-00044-5} {\bibfield  {journal}
  {\bibinfo  {journal} {The European Physical Journal D}\ }\textbf {\bibinfo
  {volume} {57}},\ \bibinfo {pages} {303} (\bibinfo {year} {2010})}\BibitemShut
  {NoStop}%
\bibitem [{\citenamefont {Press}\ \emph {et~al.}(2007)\citenamefont {Press},
  \citenamefont {Teukolsky}, \citenamefont {Vetterling},\ and\ \citenamefont
  {Flannery}}]{press2007}%
  \BibitemOpen
  \bibfield  {author} {\bibinfo {author} {\bibfnamefont {W.~H.}\ \bibnamefont
  {Press}}, \bibinfo {author} {\bibfnamefont {S.~A.}\ \bibnamefont
  {Teukolsky}}, \bibinfo {author} {\bibfnamefont {W.~T.}\ \bibnamefont
  {Vetterling}},\ and\ \bibinfo {author} {\bibfnamefont {B.~P.}\ \bibnamefont
  {Flannery}},\ }\href@noop {} {\emph {\bibinfo {title} {Numerical Recipes: The
  Art of Scientific Computing}}},\ \bibinfo {edition} {3rd}\ ed.\ (\bibinfo
  {publisher} {Cambridge University Press},\ \bibinfo {address} {Cambridge},\
  \bibinfo {year} {2007})\ \bibinfo {note} {third edition}\BibitemShut
  {NoStop}%
\bibitem [{\citenamefont {Oppenheim}\ and\ \citenamefont
  {Schafer}(2010)}]{oppenheim2010}%
  \BibitemOpen
  \bibfield  {author} {\bibinfo {author} {\bibfnamefont {A.~V.}\ \bibnamefont
  {Oppenheim}}\ and\ \bibinfo {author} {\bibfnamefont {R.~W.}\ \bibnamefont
  {Schafer}},\ }\href@noop {} {\emph {\bibinfo {title} {Discrete-Time Signal
  Processing}}},\ \bibinfo {edition} {3rd}\ ed.\ (\bibinfo  {publisher}
  {Pearson},\ \bibinfo {address} {Upper Saddle River, NJ},\ \bibinfo {year}
  {2010})\ \bibinfo {note} {third edition}\BibitemShut {NoStop}%
\end{thebibliography}
\end{document}